\documentclass[onecolumn, floatfix, preprintnumbers, eqsecnum, letterpaper, superscriptaddress, nofootinbib]{revtex4}
\usepackage{graphicx}
\usepackage{microtype}
\usepackage{amsmath}
\usepackage{amssymb}
\usepackage{subfigure}
\usepackage{hyperref}
\usepackage{url}
\usepackage{xcolor}
\usepackage{color}
\usepackage{mathrsfs}
\usepackage{calrsfs}
\usepackage{amsfonts}
\usepackage{eufrak}
\usepackage{tabularx}
\usepackage{eucal}
\usepackage{latexsym}
\usepackage{ragged2e}
\usepackage{epsfig}
\usepackage{textcomp}
\usepackage{inputenc}

\usepackage{caption}
\usepackage{subfigure}

\usepackage{marvosym}
\usepackage{ifsym}

\usepackage{lipsum}

\DeclareCaptionJustification{justified}{\leftskip=0pt \rightskip=0pt \parfillskip=0pt plus 1fil}
\definecolor{vividviolet}{rgb}{0.62, 0.0, 1.0}
\definecolor{amaranth}{rgb}{0.9, 0.17, 0.31}
\definecolor{palatinateblue}{rgb}{0.15, 0.23, 0.89}
\definecolor{brightpink}{rgb}{1.0, 0.0, 0.5}
\definecolor{cornflowerblue}{rgb}{0.39, 0.58, 0.93}
\definecolor{deepcarminepink}{rgb}{0.94, 0.19, 0.22}
\definecolor{radicalred}{rgb}{1.0, 0.21, 0.37}

\hypersetup{ linktoc=all,
	colorlinks, linkcolor={palatinateblue},
	citecolor={brightpink}, urlcolor={amaranth}
}

\graphicspath{{Images/}}

\graphicspath{{Images/}}

\def\@fnsymbol#1{\ensuremath{\ifcase#1\or \ddagger \or  $\textleaf$  \or \dagger
		\else\@ctrerr\fi}}%

\makeatother
\def\sideremark#1{\ifvmode\leavevmode\fi\vadjust{\vbox to0pt{\vss% the remark
			\hbox to 0pt{\hskip\hsize\hskip1em%                          will appear only
				\vbox{\hsize1.3cm\tiny\raggedright\pretolerance10000%          on the side
					\noindent #1\hfill}\hss}\vbox to8pt{\vfil}\vss}}}%
\def\beq{\begin{equation}}
	\def\eeq{\end{equation}}

\begin{document}
\title{Probing a regular black hole within asymptotically safe gravity via strong gravitational lensings and optical appearances}
%\author{Zonghai Li}
%\affiliation{Center for Astrophysics, School of Physics and Technology, Wuhan University, Wuhan 430072, China}
%\date{\today}
\author{Xiao-Jun
Gao}\email[email:~]{gaoxiaojun@gznu.edu.cn}
\affiliation{School of Physics and Electronic Science, Guizhou Normal University,
Guiyang 550025, People's Republic of China}
%\affiliation{College of Physics, Nanjing University of Aeronautics and Astronautics, Nanjing 211106, China}

\begin{abstract}
In this paper, we investigate the strong gravitational lensing effects and optical appearances around a spherically symmetric regular black hole, whose metric is derived from a non-singular collapsing dust ball model in asymptotically safe gravity. In this regular black hole spacetime, we obtain the analytical expression of the light deflection angle via calculating the strong field limit coefficients, and evaluate the lensing observables in strong field regime by supposing the regular black hole as the candidate of $M87^*$ and $SgrA^*$ supermassive black holes, respectively. Then, we study the optical appearances of an optically and geometrically thin spherical accretion around the regular black hole in this framework. Finally, we compare above results and show how much they differ from those obtained in the Schwarzschild black hole spacetime. We expect our results will be useful in the future to distinguish these non-singular black holes from their classical singular counterparts.

%In addition, we also in detail analyze the effects of the scale parameter $\xi$ on the strong deflection angle and the lensing observables. Finally, we study the optical appearances of an optically and geometrically thin spherical accretion around regular black hole in this framework. Comparing to the Schwarzschild black hole, we find that the photon ring of regular black hole is wider for the static accretion, but the brightness of ring is fainter, while the brightness for the infalling accretion is stronger.
\end{abstract}
%\keywords{Strong gravitational lensing, black hole shadows, asymptotically safe gravity}

\maketitle
%\textbf{Keywords:} Strong gravitational lensing; regular black hole shadow; asymptotically safe gravity

%\tableofcontents

\section{Introduction}
It is well-known black holes predicted by General Relativity (GR) \cite{Einstein1916} are one of the most fascinating  celestial objects, describing regions of spacetime with intense gravitational fields
and unique properties. More recently, there are already experiment observations manifested the existence of black holes in our universe. For example, the LIGO-Virgo collaboration has successfully detected the gravitational waves from the merger of a binary black holes \cite{LIGO:2016aoc}, and the Event Horizon Telescope (EHT) collaboration directly provided us with the first-ever images of black hole shadow in the core of the galaxy $M87^*$ \cite{EHT:2019dse,EHT:2019ths,EHT:2019ggy} and $SgrA^*$ \cite{EHT:2022wkp,EHT:2022wok}.

Theoretically, if the Kretschmann scalar is divergence as the radial coordinate $r\rightarrow 0$, the spacetime singularity \cite{Penrose:1964wq,Hawking:1970zqf}
concealed within the event horizon of black hole will occur, which results in a loss of causality and the breakdown of standard physical laws. Originally, according to cosmic censorship conjecture \cite{Penrose:1969pc}, Penrose thought that the naked singularities should be hidden beneath the event horizon of a black hole. Since then, physicists have attempted different methodologies to resolve the spacetime singularity problem. One of the most widely investigative approaches is the regular black hole models \cite{Lan:2023cvz}.
Early research idea that the spacetime singularity was replaced by a regular patch of de Sitter space \cite{Frolov:1989pf}. Bardeen implemented the first regular black hole model with the de Sitter core \cite{Bardeen1968}, which is called the Bardeen black hole. Ayon-Beato and Garcia interpreted the Bardeen black hole was a magnetic solution to Einstein equations coupled to a nonlinear electrodynamics \cite{Ayon-Beato:2000mjt}. Hitherto, a wide variety of regular black hole models have been constructed from a theoretical perspective (see, e.g., Refs. \cite{Hayward:2005gi,Bronnikov:2005gm,Burinskii:2001bq,Fan:2016hvf,Ovalle:2023ref,Mazza:2023iwv,Modesto:2010rv,Casadio:2023iqt,Lewandowski:2022zce,Carballo-Rubio:2019fnb}).

The regular black hole models have been got remarkable attentions in  asymptotic safety gravity \cite{Bonanno:2000ep,Torres:2017ygl,Eichhorn:2012va,Pawlowski:2023dda,Stashko:2024wuq,Spina:2024npx,Bonanno:2023rzk}.
In particular, recent a new model of regular black hole was presented by Bonanno et al. \cite{Bonanno:2023rzk} by generalizing an initial idea introduced Markov and Mukhanov \cite{Markov:1985py}.  It is based on the description of dynamically collapsing matter, and what is particularly interesting is that it directly predicts the formation of a regular black hole during the collapse process. According to gravity's antiscreening behavior at small distances,
the model operates under the assumption that black hole solutions observed in nature are sourced by a matter interior whose evolution is nonsingular. This mechanism is implemented starting from an effective Lagrangian that incorporates a multiplicative coupling with the matter component,  distinguishing it from other models. This regular black hole solution obtained by Bonanno et al. in  asymptotic safety gravity has been extensively investigated in \cite{Stashko:2024wuq,Spina:2024npx,Bhattacharjee:2025xcb,Mustafa:2025cou,Mannobova:2025uqf,Urmanov:2025nou,Zhao:2025sck,Turakhonov:2025ojy}. However, studies of gravitational lensing (SL) signatures in its strong field regime remain conspicuously absent, leaving a critical gap that demands detailed exploration.

The light rays pass very close to the strong field regime around a compact object, i.e., like a black hole, they will be extremely deflected. This phenomenon is usually called strong gravitational lensing (SGL). Darwin first investigated the strong deflection angle of light near the Schwarzschild black hole \cite{Darwin1961}. Latter, Virbhadra and Ellis investigated the Schwarzschild black hole lensing, and found a sequence of relativistic
images formed by light passing close to the black hole's event horizon \cite{Virbhadra:1999nm,Virbhadra:2002ju}. Following developments by Fritelli, Kling and Newman constructed the exact lens equation in the Schwarzschild black hole spacetime, and gave the analytic expressions for magnifications and time delays of relativistic images \cite{Frittelli:1999yf}. In a general static, spherically symmetric and asymptotically flat black hole spacetime, Bozza demonstrated that the light deflection angle exhibits a logarithmic divergence as the light very closes to the photon sphere \cite{Bozza:2002zj}. In this literature, Bozza also indicated that the positions and the magnifications of relativistic images around a photon sphere relate to the strong field limit coefficients \cite{Bozza:2002zj}, which carry information of black holes. Thus, if they are given by the SGL experiment observables,
we are able to identify the nature of the lensing black hole unambiguously. Recently,  Tsukamoto provided an improved strong deflection limit analysis
in a general asymptotically flat, static, spherically symmetric spacetime \cite{Tsukamoto:2016jzh}. This technique has
been extensively applied to different specific metrics, and for recent applications see \cite{Zhang:2024sgs,Wang:2024iwt,Gao:2022gbn,QiQi:2023nex,Furtado:2020puz,Nascimento:2020ime,Soares:2024rhp,Kuang:2022ojj,Kuang:2022xjp,Gao:2021lmo,Virbhadra:2024xpk,Virbhadra:2022iiy,Fu:2021fxn,Zhang:2017vap,Shipley:2019kfq}.

The black hole shadow would be formatted due to SGL around black hole \cite{Synge:1966okc}. In recent years, the related investigations of black hole shadow are the important hot topic, since its images may carry some valuable information of the spacetime geometry around black hole. Particularly, the observable features of astrophysical black hole surrounded by a luminous accretion flow are abundantly investigated. Considering a thin-disk accretion flow model, Luminet presented that the observable appearances of black hole depend on the accretion flow position and profile \cite{Luminet:1979nyg}. Falcke et al. thought that the shadow of a supermassive black hole surrounded a hot optically-thin accretion flow in the center of our Galaxy is equivalent to the GL effect by using a ray-tracing code \cite{Falcke:1999pj}. Cunha et al. investigated the lensing and the shadow of the Schwarzschild black hole surrounded by a thin and heavy accretion disk \cite{Cunha:2019hzj}. Gralla et al. studied the observable appearances of the Schwarzschild black hole with an optically/geometrically thin disk accretion flow, and shown that the lensed ring together with photon ring contribute additional observed flux to the image \cite{Gralla:2019xty}. Narayan et al. considered the spherical accretion of optically thin near the Schwarzschild black hole, and found that the shadow size and shape are independent of the accretion flow \cite{Narayan:2019imo}. By considering the spherical accretion flow, Zeng et al. analyzed influence of the quintessence dark energy and Gauss-Bonnet coupling parameter on the optical appearance of the black hole, and found that the black hole shadow depends on the physical properties of the accretion flow \cite{Zeng:2020dco,Zeng:2020vsj}. The optical appearances of black holes have been extensively studied in \cite{Zeng:2022fdm,Zeng:2024ptv,Gao:2023mjb,Zeng:2022pvb,Hu:2022lek,Saleem:2023pyx,Zeng:2021mok,Gan:2021xdl,Gan:2021pwu,Guerrero:2021ues}.

In this paper, we mainly have three goals. First, we calculate the strong deflection angle of light in the regular black hole spacetime introduced by Bonanno et al. in \cite{Bonanno:2023rzk} within asymptotically safe gravity. Second, based on the analytical expression of the strong deflection angle, we further study the observables in strong field regime related to the respective GL, and evaluate them by supposing the supermassive black holes $M87^*$ and $SgrA^*$ as the lens by this regular black hole in asymptotically safe gravity, respectively. Final, we investigate the optical appearances around the regular black hole illuminated by thin spherical accretion flows in the gravity framework.

Our paper is organized as follows: In Sec.~\ref{section2}, we first do a brief review on the static spherically symmetric regular black hole solution within asymptotically safe gravity, then following the technique introduced by Bozza \cite{Bozza:2002zj} and improved by Tsukamoto \cite{Tsukamoto:2016jzh}, we calculate the strong deflection angle around a regular black hole in asymptotically safe gravity, and evaluate the lensing observables of the supermassive black holes in this gravity framework. In Sec.~\ref{section4}, we investigate the optical features of regular black hole surrounded by an optically and geometrically thin accretion flow with spherically symmetric. Finally Sec.~\ref{section5} is for conclusions. Throughout this paper we use the geometric units with $G=c=1$ unless otherwise specified.

\section{Strong gravitational lensings by a regular black hole in asymptotically safe gravity}\label{section2}
In this section, we first make a brief review on the static spherically symmetric regular black hole solution to the corresponding field equations in asymptotically safe gravity. Then let us follow the methodology introduced by Bozza \cite{Bozza:2002zj} and improved by Tsukamoto \cite{Tsukamoto:2016jzh}, we calculate the strong deflection angle of light around the regular black hole in this gravity framework, and evaluate the lensing observables in the strong field regime by utilizing the data of the supermassive $SgrA^*$ and $M87^*$ black holes.

\subsection{The spherically symmetric regular black hole metric in asymptotically safe gravity}
By extending the ideas of Markov and Mukhanov \cite{Markov:1985py}, Bonnano et al. \cite{Bonanno:2023rzk} proposed a model of regular black hole during the gravitational collapse. The
action for the system is written as follows \cite{Bonanno:2023rzk}\footnote{In this subsection, we set $G_N\neq 1$ to better exhibit the derivation process of the regular black hole solution.}:
\begin{align}
S=\dfrac{1}{16\pi G_N}\int d^4x\sqrt{-g}[R+2\chi(\epsilon)\mathcal{L}],\label{action}
\end{align}
where $G_N$ denotes the Newton gravitational constant, $\chi(\epsilon)$ represents a multiplicative gravity-matter coupling\footnote{$\chi(\epsilon)$ has the specific property $\chi(\epsilon=0)=8\pi G_N$.}, and $\mathcal{L}$ is the matter Lagrangian. From the total variation of the action (\ref{action}), one can give the following field equations \cite{Bonanno:2023rzk}
\begin{align}
R_{\mu\nu}-\dfrac{1}{2}g_{\mu\nu}R=8\pi G(\epsilon)T_{\mu\nu}-\Lambda(\epsilon)g_{\mu\nu},\label{safefieldeq}
\end{align}
where the energy-momentum tensor $T_{\mu\nu}=[\epsilon+p(\epsilon)]u_{\mu}u_{\nu}+pg_{\mu\nu}$ is perfect fluid, among $\epsilon$ and $u_{\mu}$ represent the proper density and four-velocity of matter fluid, respectively, and the effective Newton constant $G(\epsilon)$ and cosmological
constant $\Lambda(\epsilon)$ are given by
\begin{align}
G(\epsilon)=\dfrac{\partial{\chi(\epsilon)\epsilon}}{\partial\epsilon}, \quad \Lambda(\epsilon)=-\dfrac{\partial\chi(\epsilon)}{\partial\epsilon}\epsilon^2.\label{GLambdaep}
\end{align}
The behavior of $G(\epsilon)$ as a function of the energy scale is governed by a renormalization group trajectory close
to the ultraviolet fixed point of the Asymptotic Safety program \cite{Bonanno:2019ilz,Bonanno:2021squ}, which is expressed as follows:
\begin{align}
G(\epsilon)=\dfrac{G_N}{1+\xi\epsilon},\label{Gepseq}
\end{align}
where $\xi$ is a scale parameter. The exact value of $\xi$ is unknown and
should be constrained from the observations \cite{Bonanno:2023rzk}.

As a result of the gravitational collapse of dust $(p=0)$, the spherically symmetric metric of the static exterior spacetime is derived by \cite{Bonanno:2023rzk}
\begin{align}
ds^2=-A(r)dt^2+B(r)dr^2+C(r)(d\theta^2+\sin^2\theta d\phi^2),\label{safeline}
\end{align}
with
\begin{align}
A(r)=\dfrac{1}{B(r)}=1-\dfrac{r^2}{3\xi}\log\left(1+\dfrac{6M\xi}{r^3}\right),~C(r)=r^2\label{safemetric}
\end{align}
where $M$ denotes the mass of the configuration. For details of the above derivation, see Ref. \cite{Bonanno:2023rzk}. The literature \cite{Stashko:2024wuq} has indicated that the $A(r)$ has two roots in $\xi\in(0, 0.4565M^2)$, which correspond to  the inner horizon $r^{in}_h\in(0,1.2516M)$ and  the outer horizon $r^{out}_h\in(1.2516M,2M)$, respectively. In this paper, according to this value interval of the $\xi$, we study the light deflection angle and the lensing observables of regular black hole in asymptotically safe gravity in the strong field limit.

\subsection{Strong field deflection angle of light}
For the static four-dimensional spherically symmetric spacetime in (\ref{safeline}), $A(r)\rightarrow 1$, $B(r)\rightarrow 1$ and $C(r)\rightarrow r^2$ as $r\rightarrow \infty$ in an asymptotic flat spacetime. Due to the spherically symmetric of the spacetime (\ref{safeline}), the total energy and angular momentum of the particle are constant along the trajectory, which are defined as
\begin{align}
 E\equiv-g_{\mu\nu}t^{\mu}k^{\nu}=A(r)\dot{t}, \quad L\equiv g_{\mu\nu}\phi^{\mu}k^{\nu}=C(r)\dot{\phi}, \label{ELdefine}
\end{align}
where $t^{\mu}$ and $\phi^{\mu}$ are time translational and axial Killing vectors, respectively; $k^{\nu}\equiv \dot{x}^{\nu}$ is the wave number of the particle, and $\dot{x}^{\nu}$
represents the differentiation with respect to an affine parameter of particle trajectory.

We assume that a photon starting from infinity travels on the equatorial plane ($\theta=\pi/2$ and $\dot{\theta}=0$) of a compact object, and it is deflected when passing  a closest distance $r_0$ around the compact object and then goes to infinity. The null geodesic equation is expressed as
\begin{align}
-A(r)\dot{t}^2+B(r)\dot{r}^2+C(r)\dot{\phi}^2 =0 .\label{nullgeodesic}
\end{align}
The effective potential $V_{eff}(r)$
for the motion of a photon is defined as \cite{Tsukamoto:2016jzh}
\begin{align}
V_{eff}(r)\equiv \dot{r}^2=\dfrac{L^2}{B(r)C(r)}\left(\dfrac{C(r)}{A(r)b^2}-1\right),\label{Veffdefine}
\end{align}
where $b\equiv L/E$ called the impact parameter. The equation (\ref{Veffdefine}) satisfies that $V'_{eff}(r)=0$ and $V''_{eff}(r)\leqslant 0$ and it is the largest positive root called the radius of the photon sphere $r_m$ \cite{Claudel:2000yi}.
According to (\ref{ELdefine}) and (\ref{nullgeodesic}), the trajectory of light on the equatorial plane is given by \cite{Hu:2013eya,Gao:2019pir}
\begin{align}
\dfrac{{\rm d}r}{{\rm d}\phi}=\sqrt{\frac{C(r)}{B(r)}}\sqrt{\dfrac{C(r)}{A(r)}\dfrac{1}{b^2}-1}, \label{trajectory}
\end{align}
where the impact parameter $b$ is related to the closest distance $r_0$ through the relation \cite{Keeton:2005jd,Gao:2024ejs}
\begin{align}
b=\sqrt{\dfrac{C(r_0)}{A(r_0)}}.\label{bandr0}
\end{align}
From trajectory equation (\ref{trajectory}), the deflection angle is expressed as \cite{Virbhadra:1998dy}
\begin{align}\label{deflection angle}
\hat{\alpha}(r_0)=I(r_0)-\pi,
\end{align}
where
\begin{align}
I(r_0)\equiv2\int^{\infty}_{r_0} \dfrac{\sqrt{B(r)}}{\sqrt{C(r)} \sqrt{\dfrac{C(r)}{A(r)}\dfrac{A(r_0)}{C(r_0)}-1}}{\rm d}r.\label{IntegralFunction}
\end{align}
The literature \cite{Bozza:2002zj} has demonstrated that the light deflection angle is logarithmical divergence  when $r_0$ approaches the photon sphere radius $r_m$. Recently, the work \cite{Tsukamoto:2016jzh} provided an improved of the method presented in \cite{Bozza:2002zj}.

To simplify calculation,  Tsukamoto given the definition of a variable $z$ as follows \cite{Tsukamoto:2016jzh}\footnote{Here we don't use the Bozza's definition $z=\dfrac{A(r)-A(r_0)}{1-A(r_0)}$, the reason is that the $r$ can't be expressed explicitly in terms of $z$ for the specific metric in (\ref{safemetric}).}:
\begin{align}
z=1-\dfrac{r_0}{r},\label{zdefine}
\end{align}
and then $I(r_0)$ in (\ref{IntegralFunction}) is written as
\begin{align}
I(r_0)=\int^{1}_{0}f(z,r_0)dz,\label{fzr0function}
\end{align}
where
\begin{align}
f(z,r_0)=\dfrac{2r_0}{\sqrt{\dfrac{R(z,r_0)C(z)}{B(z)}(1-z)^4}} \label{IroRz}
\end{align}
with
\begin{align} R(z,r_0)\equiv\frac{A(r_0)C(z)}{A(z)C(r_0)}-1.
\end{align}
Note that the $f(z,r_0)$ diverges when $z\rightarrow 0$, i.e., equivalent to $r\rightarrow r_0$. Therefore, the function $I(r_0)$ can  be split into a divergent part $I_D(r_0)$ and a regular part $I_R(r_0)$
\begin{align}
I_D(r_0)=&\int^{1}_{0}f_D(z,r_0){\rm d}z,\label{DivergentPart}\\
I_R(r_0)=&\int^{1}_{0}[f(z,r_0)-f_D(z,r_0)]{\rm d}z,\label{RegularPart}
\end{align}
where $f_D(z,r_0)=2r_0/\sqrt{c_1z+c_2z^2}$, among $c_1$ and $c_2$ are coefficients of the series expansion of the function beneath the square-root sign in (\ref{IroRz}), and which were given in \cite{Tsukamoto:2016jzh}.

From the above equations, one can further obtain the deflection angle in the strong field limit  \cite{Bozza:2002zj,Tsukamoto:2016jzh}
\begin{align}
\hat{\alpha}(b)=-\bar{a}\log{\left(\dfrac{b}{b_m}-1\right)}+\bar{b}+\mathcal{O}[(b-b_m)\log(b-b_m)],\label{DeAngle-b}
\end{align}
where $b_m=\lim \limits_{r \to r_m}b(r_0)$ called the critical impact parameter, and $\bar{a}$ and $\bar{b}$ are given as follows \cite{Tsukamoto:2016jzh}:
\begin{align}
\bar{a}=&\sqrt{\dfrac{2B(r_m)A(r_m)}{C''(r_m)A(r_m)-C(r_m)A''(r_m)}}\label{bara}\\
\bar{b}=&\bar{a}\log\left[r_m^2\left(\dfrac{C''(r_m)}{C(r_m)}-\dfrac{A''(r_m)}{A(r_m)}\right)\right]+b_R-\pi,\label{barb}
\end{align}
where the two prime denotes differentiation with respect to the photon sphere radius $r_m$, the regular function $I_R(r_0)$ gives a constant term $b_R$ on the photon sphere, i.e., $b_R=I_R(r_m)$. From  (\ref{DeAngle-b}), it clearly shows that the logarithmic divergence behavior of the deflection angle as $b$ approaches $b_m$.

Next, utilizing the formula of the strong deflection angle (\ref{DeAngle-b}), we can obtain the strong deflection angle $\hat{\alpha}$ of light around this regular black hole in asymptotically safe gravity. We easily obtain the radius of photon sphere from (\ref{Veffdefine})
\begin{align}
 r_m=&M+\sqrt[3]{M^3+\sqrt{9 M^2 \xi ^2-6 M^4 \xi }-3 M \xi }+\frac{M^2}{\sqrt[3]{M^3+\sqrt{9 M^2 \xi ^2-6 M^4 \xi }-3 M \xi }},\label{saferm}
\end{align}
while the critical impact parameter $b_m$ is given by
\begin{align}
 b_m=\sqrt{\dfrac{C(r_m)}{A(r_m)}}=\sqrt{\dfrac{3\xi r_m^2}{3\xi-r_m^2\log\left(1+\dfrac{6M\xi}{r_m^3}\right)}}.\label{safebm}
\end{align}
%%%%%%%%%%%%%%%%%%%%%%%%%%%%%%
\begin{figure}[h]
\centering
\includegraphics[width=2.8in]{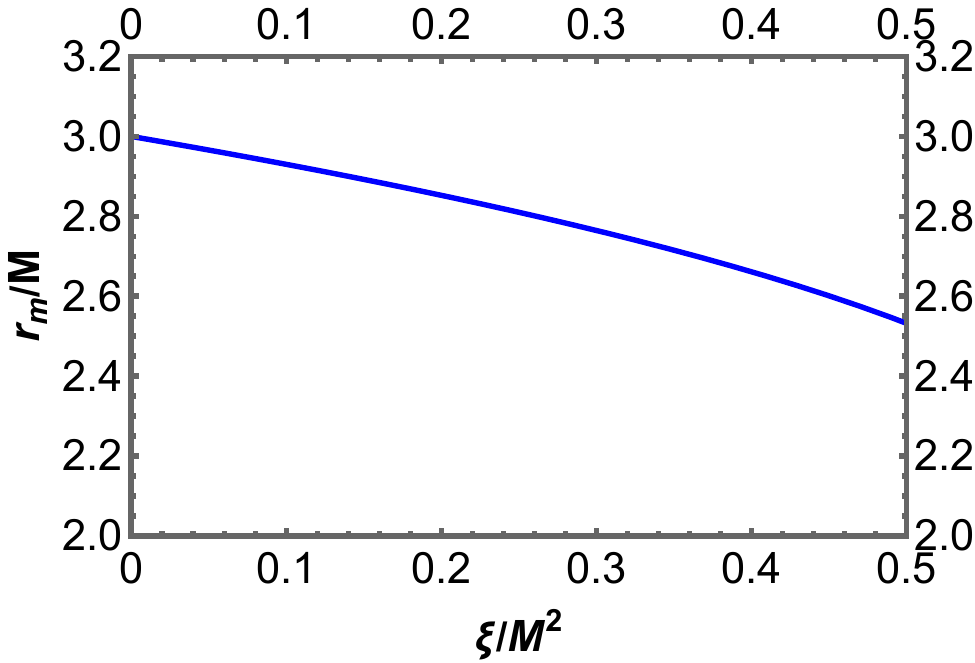}
\caption{The radius of the photon sphere $r_m$ as a function of the  scale parameter $\xi$.}
\label{rmsafe}
\end{figure}
\begin{figure}[h]
\centering
\includegraphics[width=2.8in]{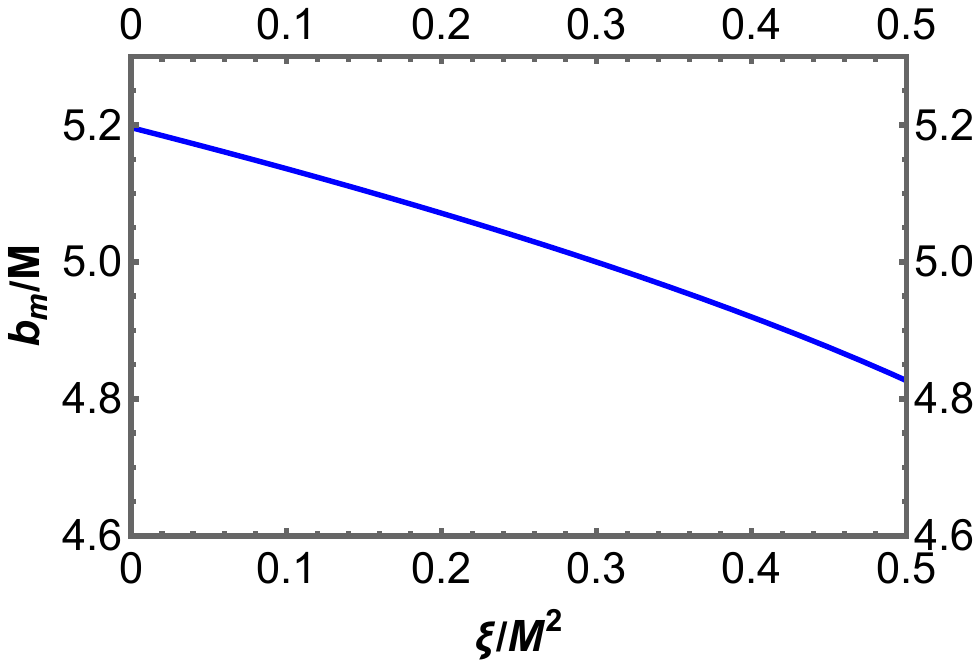}
\caption{The critical impact parameter $b_m$ as a function of the  scale parameter $\xi$.}
\label{bmsafe}
\end{figure}
%\begin{widetext}{2}
%\begin{figure}[htbp]
%\begin{minipage}[t]{0.48\linewidth}
%\centering
%\includegraphics[width=3in]{rmsafe.pdf}
%\caption{The radius of the photon sphere $r_m$ as a function of the  scale parameter $\xi$.}
%\label{rmsafe}
%\end{minipage}
%\hspace{0.1cm}
%\begin{minipage}[t]{0.48\linewidth}
%\centering
%\includegraphics[width=3in]{bmsafe}\\
%\caption{The critical impact parameter $b_m$ as a function of the  scale parameter $\xi$.}
%\label{bmsafe}
%\end{minipage}
%\end{figure}
%\end{widetext}
In the figures \ref{rmsafe} and \ref{bmsafe}, we plot the $r_m$ and the $b_m$ as a function of the $\xi$. We can observe that the $r_m$ and the $b_m$ monotonically decrease with the increasing of the $\xi$.

By substituting (\ref{safemetric}) into (\ref{bara}) and  (\ref{barb}), we easily obtain
\begin{align}
\bar{a}=&\sqrt{\dfrac{(r_m^3+6M\xi)^2}{r_m^6+6Mr_m^2(2r_m-9M)\xi+36M^2\xi^2}},\label{barasafe}\\
\bar{b}=&\bar{a}\log\left(\dfrac{6\xi[r_m^6+6Mr_m^2(2r_m-9M)\xi+36M^2\xi^2]}{(r_m^3+6M\xi)^2\left[3\xi-r_m^2\log\left(1+\dfrac{6M\xi}{r_m^3}\right)\right]}\right)+b_R-\pi,\label{barbsafe}
\end{align}
From the (\ref{RegularPart}), the constant term $b_R$ is given by
\begin{align}
b_R=I_R(r_m)=&\int^{1}_{0}[f(z,r_m)-f_D(z,r_m)]{\rm d}z\notag\\
=&2 \log \left(6 \left(2-\sqrt{3}\right)\right)+\dfrac{2}{15M^2}[4 \sqrt{3}-13\notag\\
&+\log (248832)-10 \log(\sqrt{3}+1)]\xi+\mathcal{O}(\xi^2).\label{bRsafe}
\end{align}
where we do the analytic calculation only to the first order of $\xi$. The strong field limit coefficients $\bar{a}$ and $\bar{b}$ as a function of the $\xi$ are shown in figures \ref{barasafe}-\ref{barbsafe}. Form the figure \ref{barasafe} to the figure \ref{barbsafe}, we can intuitively see that the $\bar{a}$ is continuously increasing with the $\xi$ increasingly, while the $\bar{b}$ gradually decreases.
\begin{figure}[h]
\centering
\includegraphics[width=2.8in]{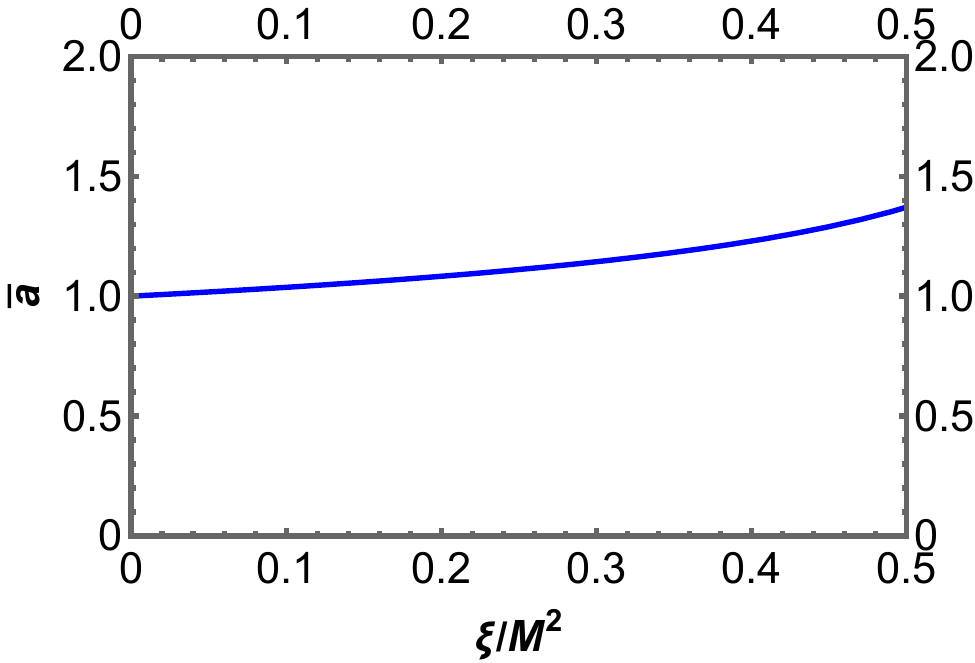}
\caption{The strong field limit coefficient $\bar{a}$ as a function of the $\xi$.}
\label{barasafe}
\end{figure}
\begin{figure}[h]
\centering
\includegraphics[width=2.8in]{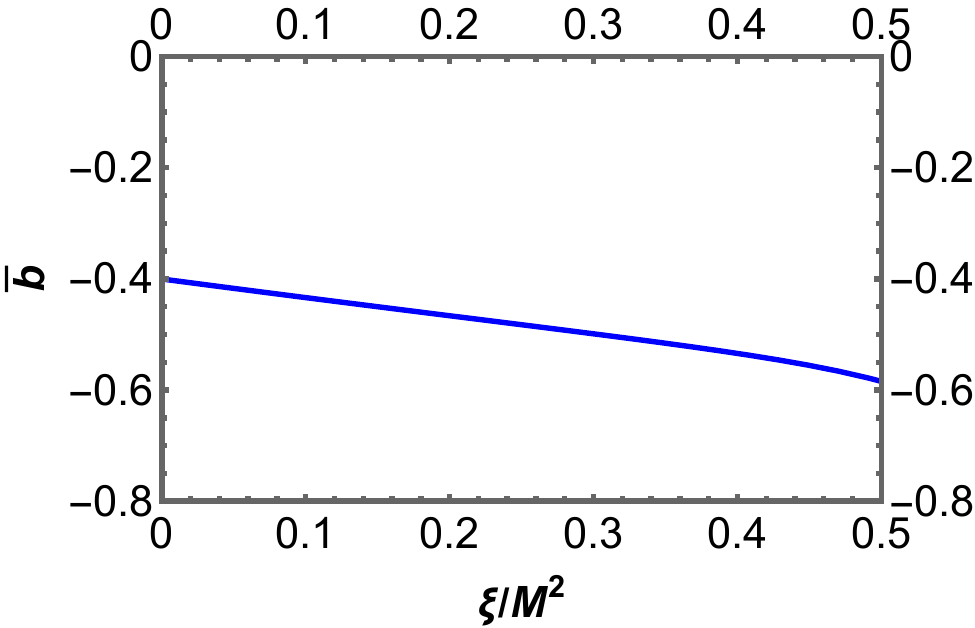}
\caption{The strong field limit coefficient $\bar{b}$ as a function of the $\xi$.}
\label{barbsafe}
\end{figure}
%\begin{widetext}{2}
%\begin{figure}[htbp]
%\begin{minipage}[t]{0.48\linewidth}
%\centering
%\includegraphics[width=3in]{barasafe.pdf}
%\caption{The strong field limit coefficient $\bar{a}$ as a function of the $\xi$.}
%\label{barasafe}
%\end{minipage}
%\hspace{0.1cm}
%\begin{minipage}[t]{0.48\linewidth}
%\centering
%\includegraphics[width=3in]{barbsafe.pdf}\\
%\caption{The strong field limit coefficient $\bar{b}$ as a function of the $\xi$.}
%\label{barbsafe}
%\end{minipage}
%\end{figure}
%\end{widetext}

\begin{figure}[h]
\centering
\includegraphics[width=2.8in]{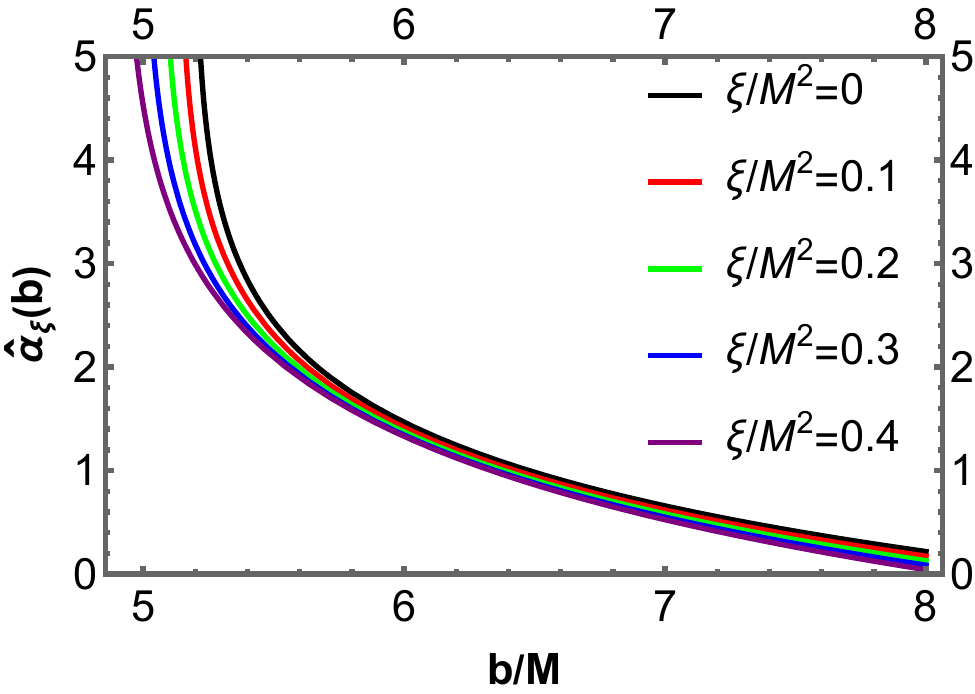}
\caption{The strong deflection angle $\hat{\alpha}_{\xi}(b)$ as a function of the impact parameter $b$ for different values of the $\xi$.}
\label{anglesafe}
\end{figure}

We further expand the strong deflection angle (\ref{DeAngle-b}) in the $\xi$ to easily read the first-order correction from the $\xi$. Putting (\ref{saferm}-\ref{bRsafe}) into (\ref{DeAngle-b}), and then expanding (\ref{DeAngle-b}) in powers of $\xi$, we finally obtain the expression of the deflection angle in the strong field limit as the consistent power-series in terms of the coefficient $\xi$
\begin{align}
\hat{\alpha}(b)=&-\left(1+\frac{\xi }{3 M^2}\right)\log\left[b-3 \sqrt{3} M+\frac{\xi }{\sqrt{3} M}\right]\notag\\
&+\log \left[648 \left(7 \sqrt{3}-12\right) M\right]-\pi+\dfrac{1}{90M^2}[30 \log (M)\notag\\
&+48 \sqrt{3}-206+150 \log (2)+135 \log (3)\notag\\
&-120 \log(\sqrt{3}+1)]\xi+(\text{terms of order} > 1).\label{finalldefangle}
\end{align}
The above result can reduce to that of the Schwarzschild black hole case in \cite{Bozza:2002zj,Tsukamoto:2016jzh} when the $\xi=0$. We have plotted the strong deflection angle of light (\ref{finalldefangle}) as a function of the impact parameter $b$ for different values of the $\xi$ in figure \ref{anglesafe}. It is found that a larger the $\xi$ leads to a smaller the deflection angle at the same impact parameter $b$.

\subsection{Observables in strong field regime}
\begin{figure}[h]
\centering
\includegraphics[width=3.2in]{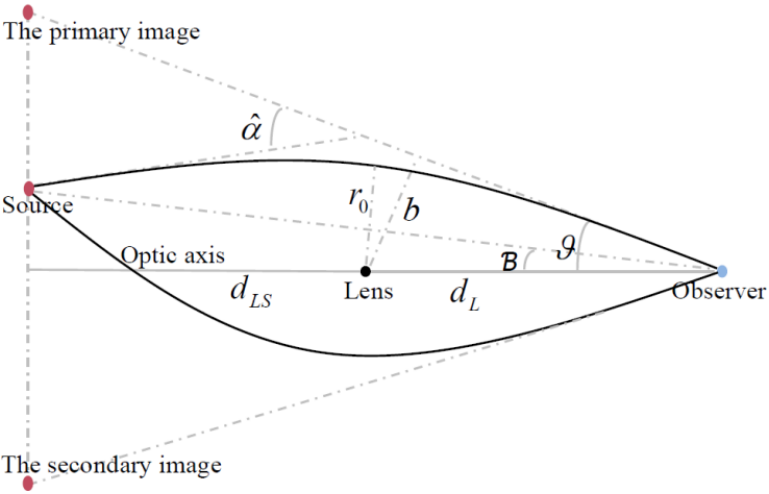}
\caption{The schematic diagram of light bending and GL \cite{Gao:2019pir}.}
\label{lenspicturenew}
\end{figure}

We further derive some lensing observables based on the deflection angle in the strong field limit. From the figure \ref{lenspicturenew}, we represent the angular position of the source by $\mathcal{B}$, the angular position of the image by $\vartheta$, and the light deflection angle by $\hat{\alpha}$. The relation expression between them is given as follows \cite{Bozza:2008ev}:
\begin{align}
d_S\tan{\mathcal{B}}=\dfrac{d_L\sin{\vartheta}-d_{LS}\sin{(\hat{\alpha}-\vartheta)}}{\cos{(\hat{\alpha}-\vartheta)}},\label{lenseqautionnew}
\end{align}
where $d_L$ denotes the distance between the observer and the lens; $d_{LS}$ is the distance from the lens to the source plane; the distance from the observer to the source plane is expressed by $d_S=d_{L}+d_{LS}$.

Following the literature \cite{Bozza:2002zj}, we assume that the source is almost perfectly aligned along the optic axis. In the limit case, $\mathcal{B}, \vartheta \ll 1$, and $\hat{\alpha}=2n\pi+\Delta\hat{\alpha}_n$, where $\Delta\hat{\alpha}_n\ll 1$ is the offset of the deflection angle. Therefore, the (\ref{lenseqautionnew}) is simplified by
\begin{align}
\mathcal{B}=\vartheta-\dfrac{d_{LS}}{d_S}\Delta\hat{\alpha}_n.\label{lensequationnew2}
\end{align}
From the figure \ref{lenspicturenew} and the above condition, one can obtain $b\simeq\vartheta d_L$. Thus, the strong deflection angle (\ref{DeAngle-b})  is rewritten as
\begin{align}
\hat{\alpha}(\vartheta)=-\bar{a}\log{\left(\dfrac{\vartheta d_L}{b_m}-1\right)}+\bar{b}\label{angledefnew}.
\end{align}
To obtain $\Delta\hat{\alpha}_n$, $\hat{\alpha}(\vartheta)$ is expanded in $\vartheta=\vartheta^0_n$, and then one can give
\begin{align}
\Delta\hat{\alpha}_n=\dfrac{\partial \hat{\alpha}(\vartheta)}{\partial \vartheta}\Big|_{\vartheta=\vartheta^0_n}(\vartheta-\vartheta^0_n),\label{Deltaalpha}
\end{align}
where $\hat{\alpha}(\vartheta^0_n)=2n\pi$. Evaluating (\ref{angledefnew}) in $\vartheta=\vartheta^0_n$, gives
\begin{align}
\vartheta^0_n=\dfrac{b_m}{d_L}(1+e_n) \quad\text{with} \quad e_n=e^{\bar{b}-2n\pi}.\label{vartheta0n}
\end{align}
Substituting (\ref{vartheta0n}) into (\ref{Deltaalpha}), the $\Delta\hat{\alpha}_n$ is given by
\begin{align}
\Delta\hat{\alpha}_n=-\dfrac{\bar{a}d_L}{b_me_n}(\vartheta-\vartheta^0_n).\label{Deltaalphasolu}
\end{align}

By using (\ref{Deltaalphasolu}) and noting that $b_m/d_L\ll 1$, and then solving the lens equation (\ref{lensequationnew2}),  the angular position of the $n^{th}$ relativistic image
\begin{align}
\vartheta_n=\vartheta^0_n+\dfrac{b_me_n}{\bar{a}}\dfrac{d_S}{d_Ld_{LS}}(\mathcal{B}-\vartheta^0_n).\label{varthetan}
\end{align}
Although the gravitational lensing has conservative surface brightness, it changes the appearance of the solid angle of the source. The magnification is the ratio of the solid angles subtended by the $n$-th image and the source, i.e., $\mu_n=\Big|\dfrac{\mathcal{B}}{\vartheta}\dfrac{\partial\mathcal{B}}{\partial\vartheta}\Big|^{-1}_{\vartheta=\vartheta^0_n}$. Then, using the (\ref{lensequationnew2}) and  recalling that (\ref{Deltaalphasolu}), we further obtain
\begin{align}
\mu_n=\dfrac{e_n(1+e_n)}{\bar{a}\mathcal{B}}\dfrac{d_S}{d_{LS}}\left(\dfrac{b_m}{d_L}\right)^2.\label{mun}
\end{align}
From the (\ref{mun}), the $\mu_n$ decreases monotonically when the $n$ gradually increases, and the luminosity of the relativistic images is very weak due to the presence of the $(b_m/d_L)^2$. However, if the $\mathcal{B}=0$, the source, lens and observer are aligned along the same
axis, which will lead to the images have large brightness. Without losing generality, one assumes that only the outermost image $\vartheta_1$ is regarded as a single image, while all the remaining ones together are encompassed in $\vartheta_{\infty}$, which represents the asymptotic position approached by a set of images. Then one can define the following three observables of the relativistic images, which read as \cite{Bozza:2002zj}:
\begin{align}
\vartheta_{\infty}=&\dfrac{b_m}{d_L},\label{varthetainf}\\
s=&\vartheta_1-\vartheta_{\infty}\simeq\vartheta_{\infty}e^{\dfrac{\bar{b}-2\pi}{\bar{a}}},\label{angsepa}\\
r_{mag}=&\dfrac{\mu_1}{\sum\limits^{\infty}_{n = 2}\mu_n}\simeq\dfrac{5\pi}{\bar{a}\log(10)},\label{relativemag}
\end{align}
where $s$ is the angular separation between the outermost and asymptotic relativistic images, and $r_{mag}$ is the relative magnification of the outermost relativistic
image. From (\ref{varthetainf}) to (\ref{relativemag}), it obviously sees that three observables of the relativistic images are determined by the strong field limit coefficients $\bar{a}, \bar{b}$, and the critical impact parameter $b_m$. Inversely, if we can successful measure the above lensing observables from experiment, which would be helpful for us to distinguish the nature of black holes or lens.

In addition, multiple relativistic images of the source are formed when the deflection angle is more than $2\pi$. Consequently, the travel time in different light paths corresponding to different images will be theoretically distinguishable, resulting in a another important observable in strong field lensing, i.e., the time delay. The time delay
between $i-th$ and $j-th$ images could be approximatively written as \cite{Bozza:2003cp}
\begin{align}
\Delta T^s_{i,j}=&2\pi(i-j)\dfrac{\tilde{a}}{\bar{a}}+2\sqrt{\dfrac{B(r_m)b_m}{A(r_m)\hat{c}}}e^{\dfrac{\bar{b}}{2\bar{a}}}\left(e^{-\dfrac{2\pi j\mp\mathcal{B}}{2\bar{a}}}-e^{-\dfrac{2\pi i\mp\mathcal{B}}{2\bar{a}}}\right)
\end{align}
for the images on same side of the lens, while
\begin{align}
\Delta T^o_{i,j}=&2\pi((i-j)-2\mathcal{B})\dfrac{\tilde{a}}{\bar{a}}+2\sqrt{\dfrac{B(r_m)b_m}{A(r_m)\hat{c}}}e^{\dfrac{\bar{b}}{2\bar{a}}}\left(e^{-\dfrac{2\pi j-\mathcal{B}}{2\bar{a}}}-e^{-\dfrac{2\pi i+\mathcal{B}}{2\bar{a}}}\right)
\end{align}
for the images on opposite side of the lens, with $\bar{a}$, $\tilde{a}$, $\bar{b}$ and $\tilde{c}$ given in \cite{Tsukamoto:2016jzh,Bozza:2003cp}. For both cases, the literature \cite{Bozza:2003cp} has demonstrated that dominant contribution of the time delay comes from the first term. Therefore, the time delay between the first and second relativistic images on same side of the lens is given as \cite{Bozza:2003cp}
\begin{align}
\Delta T^s_{2,1}=2\pi\vartheta_{\infty}d_L.\label{timedelay}
\end{align}

\subsubsection{Evaluating the observables by $M87^*$ and $SgrA^*$
supermassive black holes}
To perform the analysis of the observables, we assume that the supermassive $M87^*$ and $SgrA^*$ black
holes are regarded as the lens by the regular black hole within asymptotically safe gravity, and then evaluate and compare the lensing observables in the strong field limit with those of Schwarzschild black holes. To accomplish the aim, we will apply the realistic mass and distance of the lens, i.e., $M=4.0\times 10^6 M_{\odot}$ and $d_L=8.35~{\rm Kpc}$ for $SgrA^*$ \cite{EHT:2022wkp,EHT:2022wok}, while $M=6.5\times 10^9 M_{\odot}$ and $d_L=16.8~{\rm Mpc}$ for $M87^*$ \cite{EHT:2019dse,EHT:2019ths,EHT:2019ggy}.

\begin{figure}[htbp]
\centering
\includegraphics[scale=0.55]{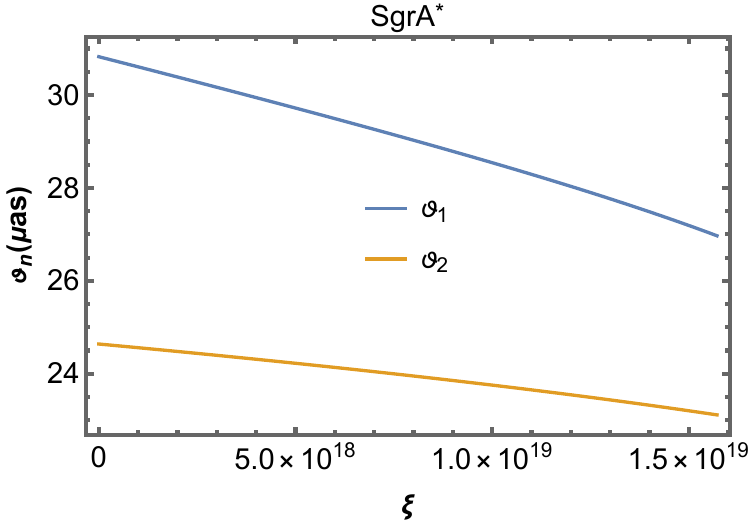}
\includegraphics[scale=0.55]{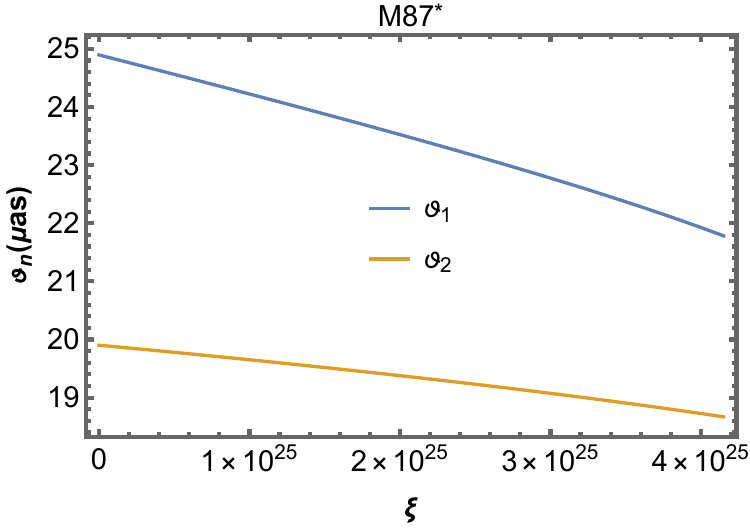}
\caption{The angular positions ($\vartheta_1$and $\vartheta_2$) of the first and second relativistic
images as a function of the $\xi$ with $\mathcal{B}=10^2$. The left panel is for $SgrA^*$ supermassive black hole while the right panel is for  $M87^*$ .}
\label{thetansafexi}
\end{figure}

\begin{figure}[!h]
\centering
\includegraphics[scale=0.55]{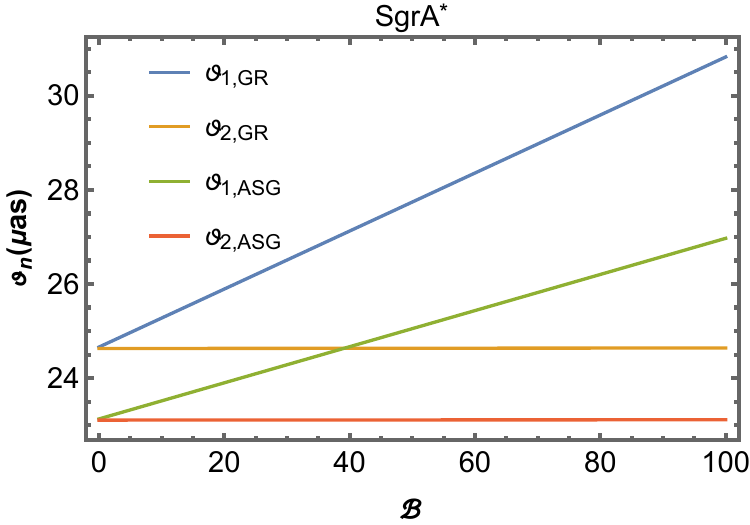}
\includegraphics[scale=0.55]{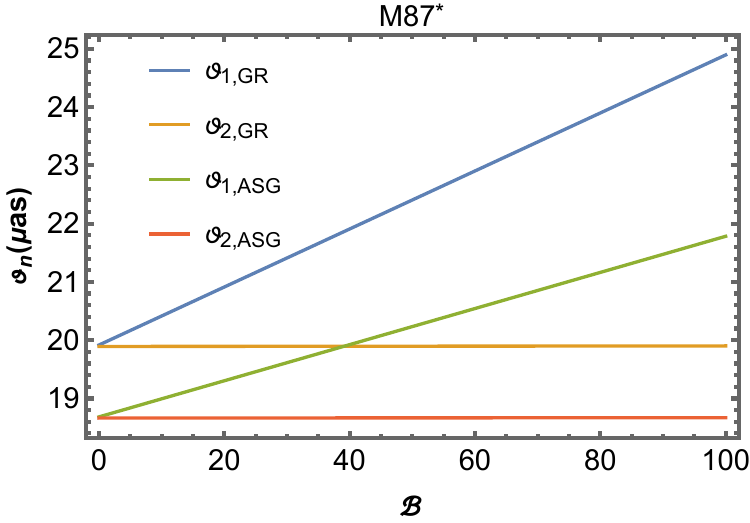}
\caption{The angular positions ($\vartheta_1$and $\vartheta_2$) of the first and second relativistic
images as a function of the $\mathcal{B}$ with $\xi=0.45M^2$. The left panel is for $SgrA^*$ supermassive black hole while the right panel is for  $M87^*$ .}
\label{thetansafebeta}
\end{figure}

By taking $d_{LS}=d_S/2$ and using above data, we have plotted that the angular positions ($\vartheta_1$ and $\vartheta_2 $) of the first and second relativistic images as a function of the $\xi$ or the $\mathcal{B}$ via the (\ref{varthetan}) in figures \ref{thetansafexi}-\ref{thetansafebeta}.
It obviously sees that $\vartheta_1$ and $\vartheta_2 $ decrease as the value of the $\xi$ increases, but the position of source $\mathcal{B}$ makes the $\vartheta_1$ larger, and slightly affects the $\vartheta_2$.
%%%%%%%%%%%%%%%%%%%%%%%%%%%%%
%%%%%%%%%%%%%%%%%%%%%%%%%%%%%
%%%%%%%%%%%%%%%%%%%%%%%%%%
\begin{figure}[!h]
\centering
\includegraphics[scale=0.55]{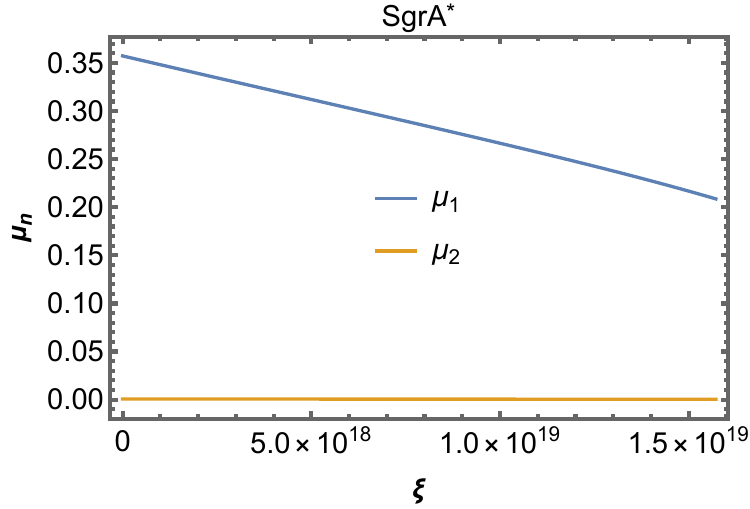}
\includegraphics[scale=0.55]{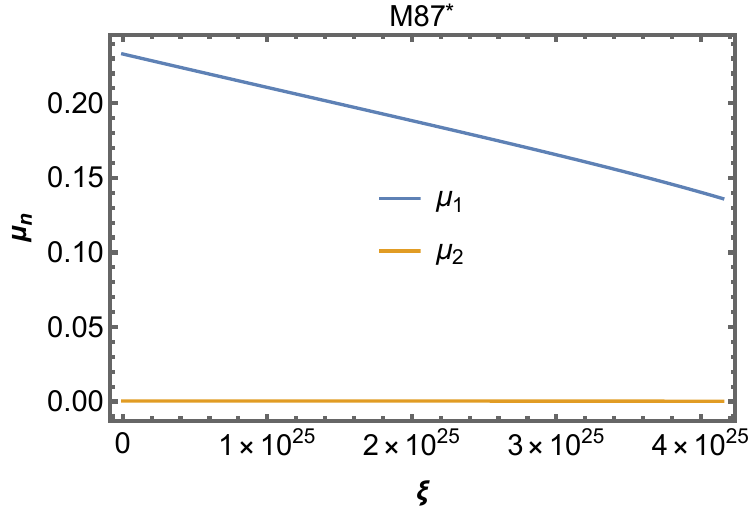}
\caption{The
magnifications ($\mu_1$and $\mu_2$) of the first and second relativistic
images as a function of the $\xi$ with $\mathcal{B}=10^2$. The left panel is for $SgrA^*$ supermassive black hole while the right panel is for  $M87^*$ .}
\label{munsafexi}
\end{figure}
%%%%%%%%%%%%%%%%%%%%%%%%%%%%%
\begin{figure}[!h]
\centering
\includegraphics[scale=0.55]{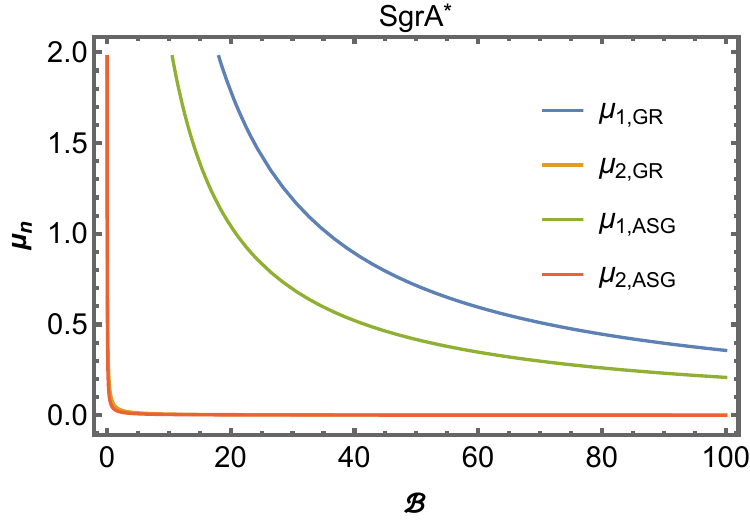}
\includegraphics[scale=0.55]{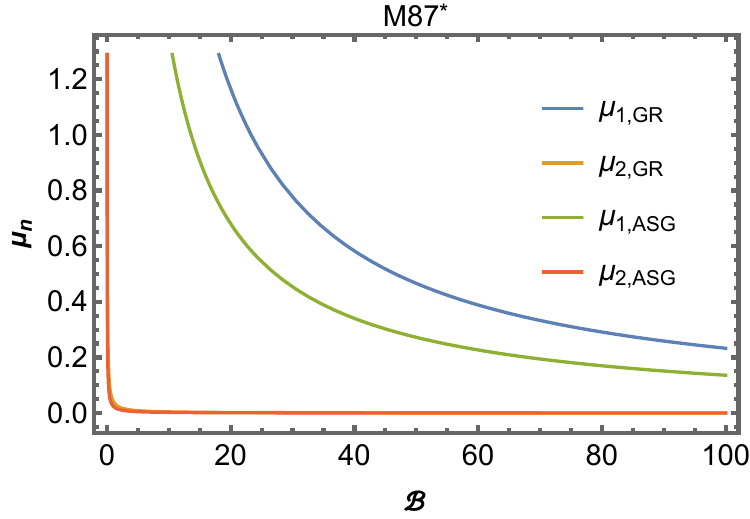}
\caption{The
magnifications ($\mu_1$and $\mu_2$) of the first and second relativistic
images as a function of the $\mathcal{B}$ with $\xi=0.45M^2$. The left panel is for $SgrA^*$ supermassive black hole while the right panel is for  $M87^*$ .}
\label{munsafebeta}
\end{figure}
%%%%%%%%%%%%%%%%%%%%%%%%%
\begin{figure}[!h]
\centering
\includegraphics[scale=0.55]{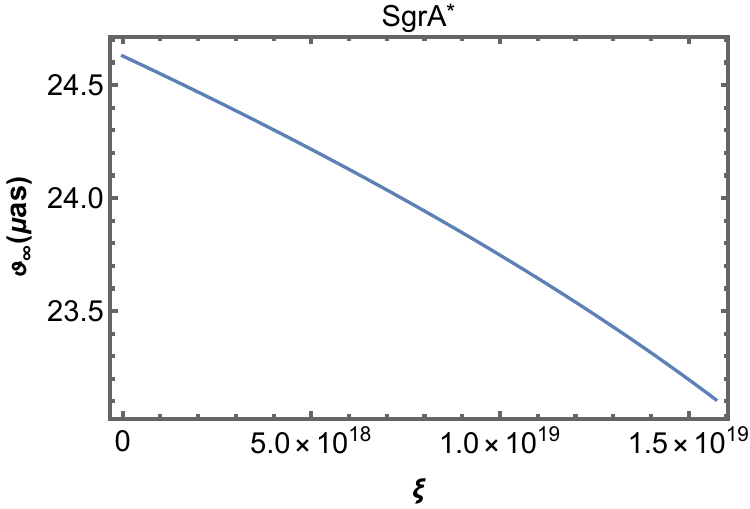}
\includegraphics[scale=0.55]{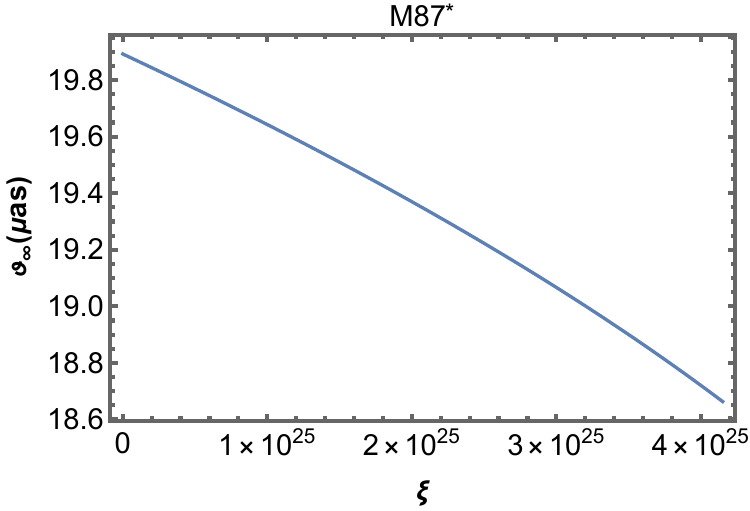}
\includegraphics[scale=0.55]{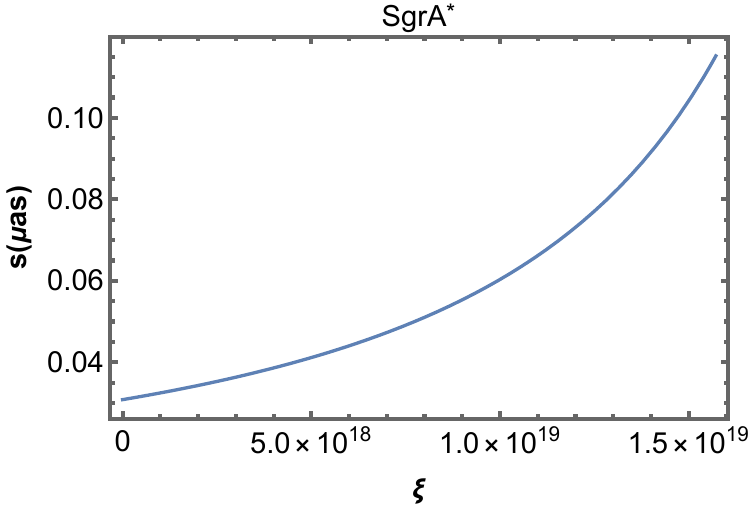}
\includegraphics[scale=0.55]{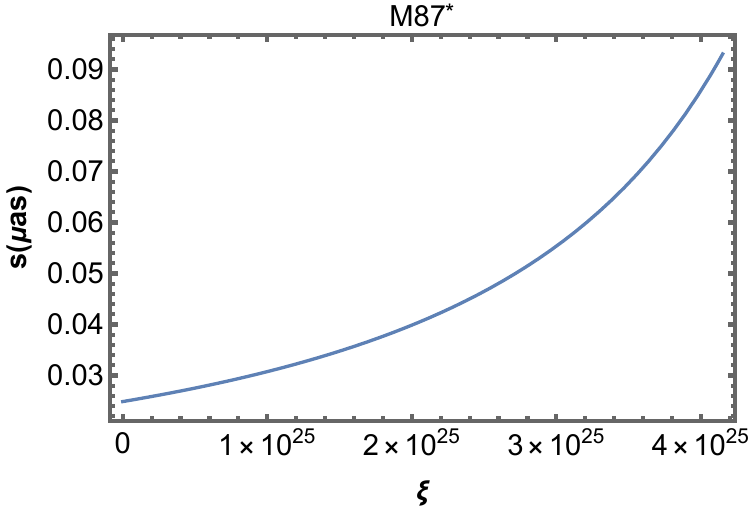}
\includegraphics[scale=0.55]{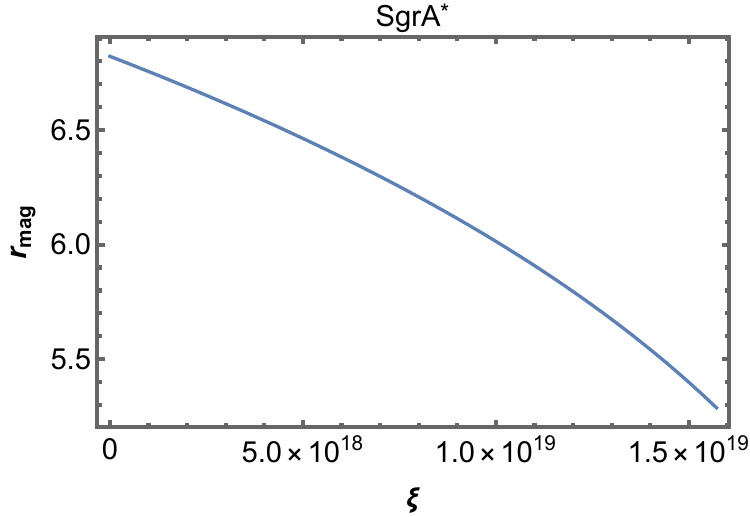}
\includegraphics[scale=0.55]{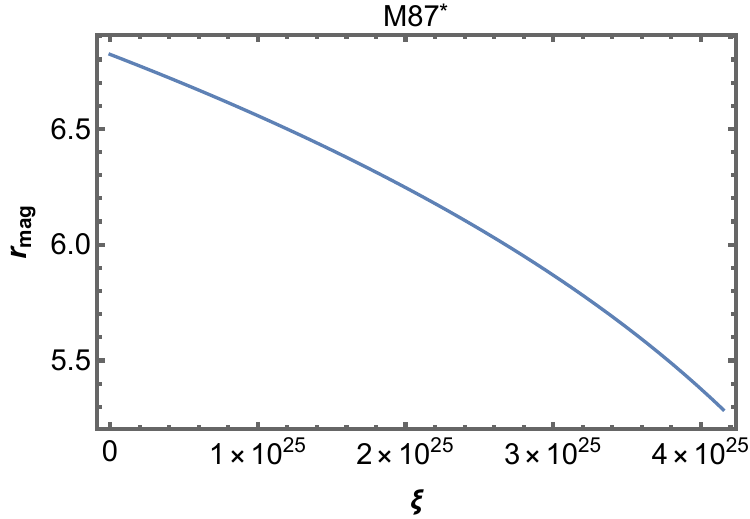}
\caption{Behaviour of strong lensing observables $\vartheta_{\infty}$, $s$ and $r_{mag}$ with the parameter $\xi$ by taking supermassive black holes $Sgr^*$
(First column) and $M87^*$ (Second column) as the regular black hole within asymptotically safe gravity, respectively.}
\label{thetasrmag}
\end{figure}
%%%%%%%%%%%%%%%%%%%%%%%%%
\begin{figure}[!h]
\centering
\includegraphics[scale=0.55]{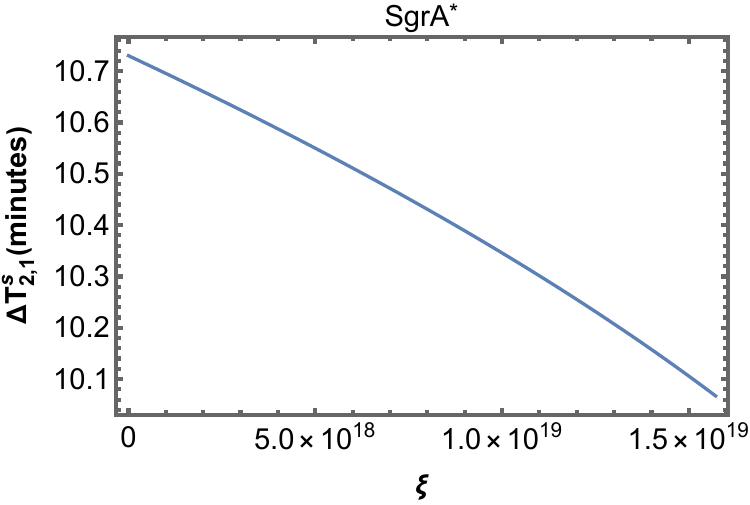}
\includegraphics[scale=0.55]{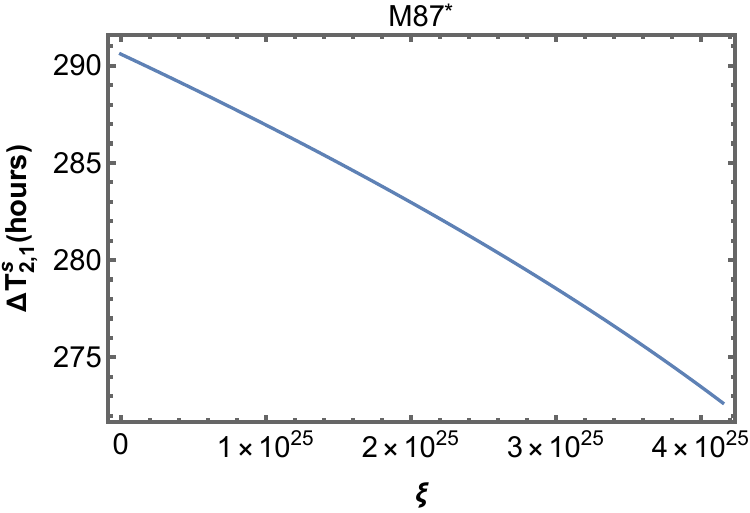}
\caption{The time delay $\Delta T^s_{2,1}$ between the first image and the second image as a function of the $\xi$ for regular black hole within asymptotically safe gravity as the $M87^*$ and $SgrA^*$ supermassive black hole, respectively.}
\label{M87Sgrtimedelay}
\end{figure}
%%%%%%%%%%%%%%%%%%%%%%%%%%%%%%
The figures \ref{thetansafexi}-\ref{thetansafebeta} show that the supermassive black holes with the scale parameter $\xi$ have the smaller angular positions in comparison with the Schwarzschild black hole cases. Next, utilizing the (\ref{mun}), the corresponding
magnification ($\mu_1$ and $\mu_2$) of the first and second order images as a function of the $\xi$ or the $\mathcal{B}$ have be shown in figures \ref{munsafexi}-\ref{munsafebeta}. We see that as
the value of the $\xi$ or the $\mathcal{B}$ increases, the $\mu_1$ decreases, while they hardly affect for the magnification of higher order images. In addition, the first order images
for the regular black hole in asymptotically safe gravity is highly magnified than the second order images. Meanwhile, we find that the images of the regular black hole in asymptotically safe gravity are darker than those of Schwarzschild black hole for the first-order images. With the same parameters, both the angular positions and the relative
magnifications of the regular black hole and their deviations from GR for
$SgrA^*$ are larger than those for $M87^*$, making it is easier to
detect in $SgrA^*$.

The characteristic observables including the position of the innermost image $\vartheta_{\infty}$, the angular separation $s$ and the relative magnification $r_{mag}$ of the outermost image given by
(\ref{varthetainf})-(\ref{relativemag}), as functions of the scale parameter $\xi$ for supermassive black holes are depicted in figure \ref{thetasrmag}. From the figure \ref{thetasrmag}, when the $\xi$ gradually increases, the $\vartheta_{\infty}$ and $r_{mag}$ monotonously descend in the region of the $\xi$, while the $s$ continuously increases. Comparison with the Schwarzschild black hole cases, the $\vartheta_{\infty}$ and $r_{mag}$ with the $\xi$ are smaller, while the $s$ is larger. In addition, the both observables $\vartheta_{\infty}$ and $r_{mag}$ are much larger than the $s$. It implies that the position of the innermost image and the relative magnification of the outermost image can better be observed than the angular separation. Moreover, by comparing the first column plots and second column plots, we find that the characteristic observables in the regular black holes and their deviation from GR are more profound for the $SgrA^*$ than $M87^*$.

Finally, we study the time delay between the first image and the second image for regular black hole within asymptotically safe gravity as the $M87^*$ and $SgrA^*$ supermassive black hole, respectively. According to the (\ref{timedelay}), the time delay as a function of the $\xi$ has been shown in figure \ref{M87Sgrtimedelay}. It is clearly seen that the scale parameter $\xi$ shortens the time delay.  In addition, the time delay for the $M87^*$ can be hundreds hours, which is much longer than the several minutes for the $SgrA^*$. Consequently, we have chance to observe such higher time delays within reasonable times of exposure for the $M87^*$ in comparison with the $SgrA^*$.

\section{The optical appearances of a regular black hole illuminated by different spherical accretion flows in asymptotically safe gravity}\label{section4}
The substances with extremely small the angular momentum in the Universe will flow radially to the black hole and form spherically symmetric accretion \cite{Yuan:2014gma}. Thus, in this section, we consider that the spherically symmetric accretion flows around black hole are the optically and geometrically thin, and then investigate the optical appearances of the regular black hole illuminated by different spherical accretion flows in asymptotically safe gravity. For a distant observer, the observed specific intensity (measured in ${\rm erg}~ {\rm s}^{-1}~ {\rm cm}^{-2}~ {\rm str}^{-1}~ {\rm Hz}^{-1}$) at $r=\infty$ radiated by the accretion flow can be obtained by integrating the specific emissivity along the photon path $\gamma$ \cite{Jaroszynski:1997bw}
\begin{align}
I(\nu_o)=\int_{\gamma} g^{3}j_e(\nu_{e})dl_{prop},\label{obsintenEq}
\end{align}
where the red-shift factor $g$ is the ratio between the observed photon frequency $\nu_o$ and emitted photon frequency $\nu_e$, i.e., $g\equiv \nu_o/\nu_e$; $j_e(\nu_{e})$ represents the emissivity per unit volume in the rest frame of the emitter, which will be set $j_e(\nu_{e})\propto \delta(\nu_r-\nu_e)/r^2$ in this work, among $\nu_r$ denotes the emitter's rest-frame frequency; $dl_{prop}$ is the infinitesimal proper length. Integrating (\ref{obsintenEq}) over all the observed frequencies, the total observed intensity is
given by
\begin{align}
I_{obs}=\int_{\nu_o} I(\nu_o)d\nu_o=\int_{\nu_e}\int_{\gamma} g^{4}j(\nu_{e})dl_{prop}d\nu_{e}.\label{totalintenstate}
\end{align}
Next, we consider that the spherical accretion is static and radially free-infalling, respectively, and then study the influence of the scale parameter $\xi$ on the observable appearances of regular black hole.

\subsection{Static spherical accretions}\label{subsection}
The static spherical accretions distributed outside the event horizon of the regular black hole are firstly considered. For this case, the red-shift factor $g$ is equal to $f(r)^{1/2}$, and the proper length is expressed as
\begin{align}
dl_{prop}=\sqrt{\dfrac{1}{f(r)}+r^2\left(\dfrac{d\phi}{dr}\right)^2}dr.\label{properlengthEq}
\end{align}
Finally, we can obtain the total photon intensity measured by the distant
observer as follows:
\begin{align}
I_{obs}=\int_{\gamma}\dfrac{f(r)^{2}}{r^2}\sqrt{\dfrac{1}{f(r)}+\dfrac{b^2}{r^2-b^2f(r)}}dr.\label{stotalinten}
\end{align}

\begin{figure}[!h]
\centering
\includegraphics[scale=0.50]{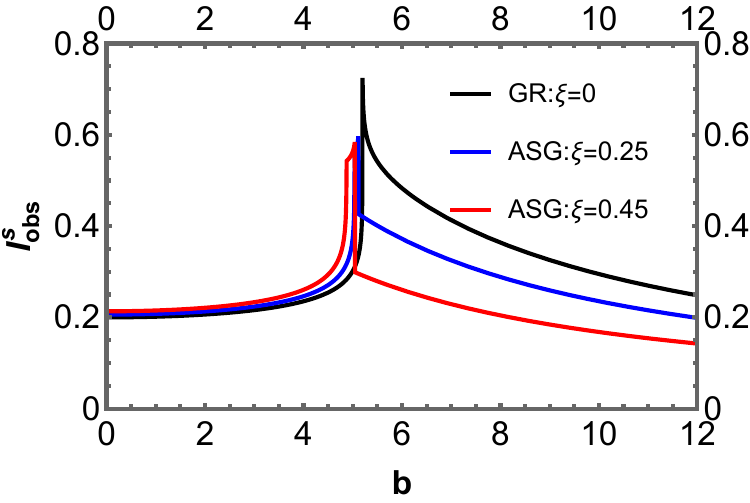}
\caption{The total observed intensities $I^s_{obs}$ as a function of the impact parameter $b$ with a static spherical accretion flow for $M=1$.}
\label{stateIb}
\end{figure}

\begin{figure}[!h]
\centering
\includegraphics[scale=0.46]{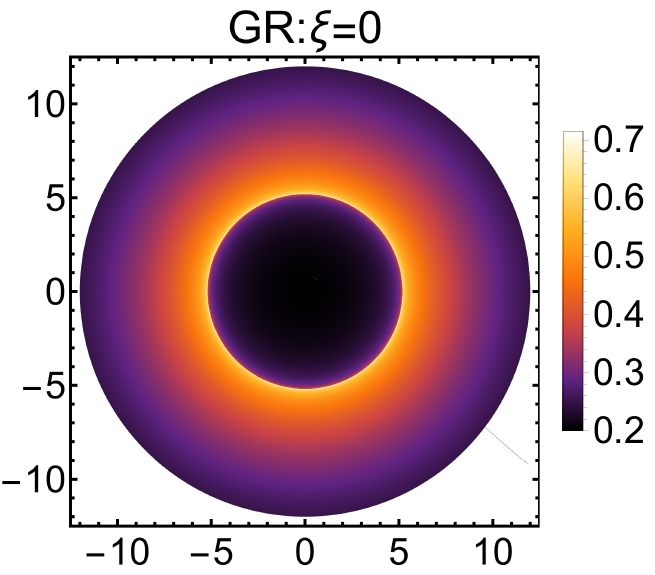}
\includegraphics[scale=0.46]{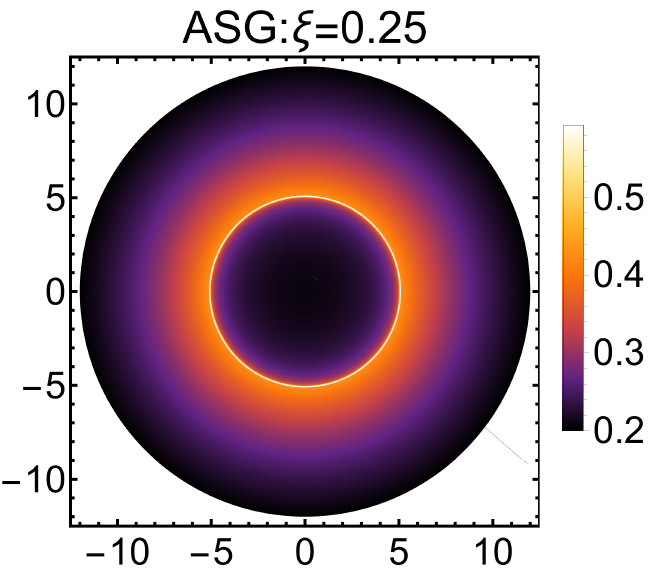}
\includegraphics[scale=0.46]{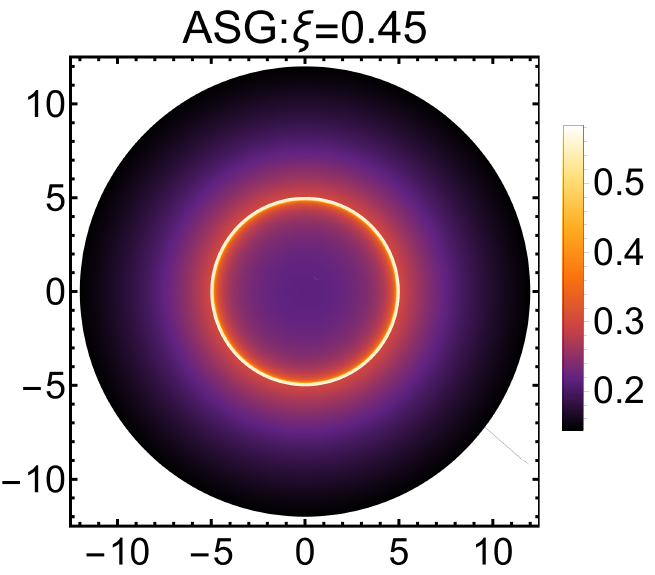}
\caption{Two-dimensional images of shadows and photon rings of regular black holes with a static spherical accretion flow for $M=1$.}
\label{stateintenfig}
\end{figure}

We have plotted the total observed intensity $I^s_{obs}$ as a function of the impact parameter $b$ in figure \ref{stateIb}, which reveals that the peak values of observed intensity at the $b=b_m$ and the shadow sizes decrease with the increasing of parameter $\xi$ by the finite numerical limitations. It also shows that the maximum brightness of regular black hole image in asymptotically safe gravity is lower than that of the Schwarzschild case. However, the larger scale parameter $\xi$ results in the wider region of photon ring in comparison with the Schwarzschild black hole. In addition, the figure \ref{stateintenfig} depicts the two-dimensional images of regular black hole illuminated by the static spherical accretion flows for equatorial observers. The faint illuminating region in the center represents the black hole shadow and the surrounding is regarded as the spherical accretion flow. Comparing to the Schwarzschild black hole, the regular black hole has smaller and fainter photon ring but larger interval, which are consistent with the results reflected in figure \ref{stateIb}.

\subsection{Radially infalling spherical accretions}
The most of matters are dynamical in realistic astrophysical settings, we then consider that the black hole is surrounded by the radially infalling freely spherical accretion flows. In the scenario, the redshift factor $g$ should be rewritten as \cite{Bambi:2013nla}
\begin{align}
g=\dfrac{k_\mu u^{\mu}_o}{k_\nu u^{\nu}_{e}},\label{infalingfac}
\end{align}
where $k_\mu$ denotes the four-momentum of photon emitted from accretion matter, the corresponding components are given by \cite{Bambi:2013nla,Wang:2023vcv}
\begin{align}
k_t=\dfrac{1}{b},k_r=\pm\sqrt{\dfrac{1}{b^2f(r)^2}-\dfrac{1}{f(r)r^2}},\label{ktkr}
\end{align}
among $\pm$ in $k_r$ describe the photons approaching $(+)$ or away $(-)$ from the black hole; $u^{\mu}_o=(1,0,0,0)$ is the four velocity of an stationary distant observer; $u^{\nu}_{e}=(u^{t}_{e},u^{r}_{e},0,0)$ is the four-velocity of the accretion flow, which read \cite{Bambi:2013nla,Wang:2023vcv}
\begin{align}
u^{t}_{e}=\dfrac{1}{f(r)},~~u^{r}_{e}=-\sqrt{1-f(r)},~~u^{\theta}_{e}=u^{i \phi}_{e}=0.\label{utrthetaphi}
\end{align}
Therefore, the redshift factor $g$ can be expressed as
\begin{align}
g=\dfrac{k_t}{k_t u^t_e+k_r u^r_e}.\label{gnew}
\end{align}
Meanwhile, the proper length is defined as \cite{Bambi:2013nla}
\begin{align}
dl_{prop}=k_\sigma u^{\sigma}_{e}d\lambda=\dfrac{k_t}{g^|k_r|}dr.\label{infallingprop}
\end{align}
We finally obtain the total observed intensity (\ref{totalintenstate}) for the radially
infalling accretion flow as follows:
\begin{align}
I_{obs}=\int_{\gamma}\dfrac{g^{3}}{r^2}\dfrac{1}{\sqrt{\dfrac{1}{f(r)}\left(\dfrac{1}{f(r)}-\dfrac{b^2}{r^2}\right)}}dr.\label{infallinginten}
\end{align}

\begin{figure}[htbp]
\centering
\includegraphics[scale=0.50]{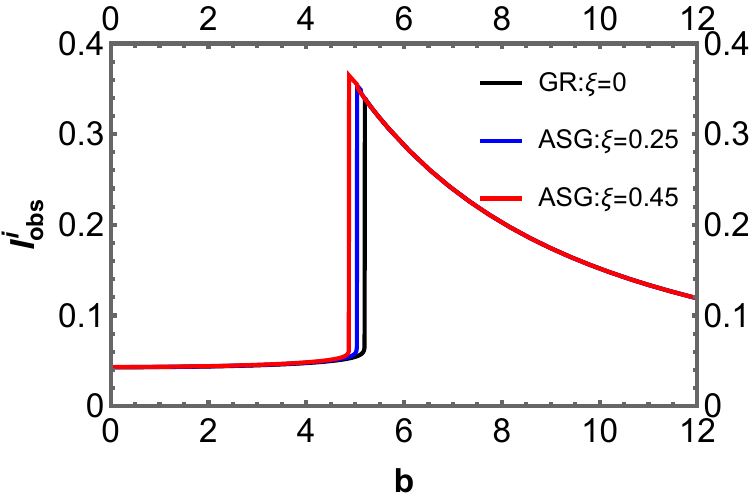}
\caption{The observed specific intensities $I^i_{obs}$ as a function of the impact parameter $b$ with an infalling spherical accretion flow for $M=1$.}
\label{infallingIb}
\end{figure}

\begin{figure}[!h]
\centering
\includegraphics[scale=0.48]{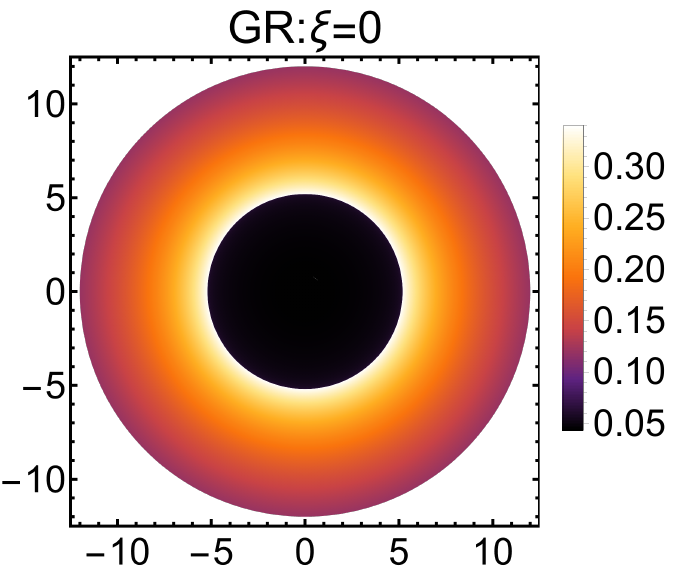}
\includegraphics[scale=0.48]{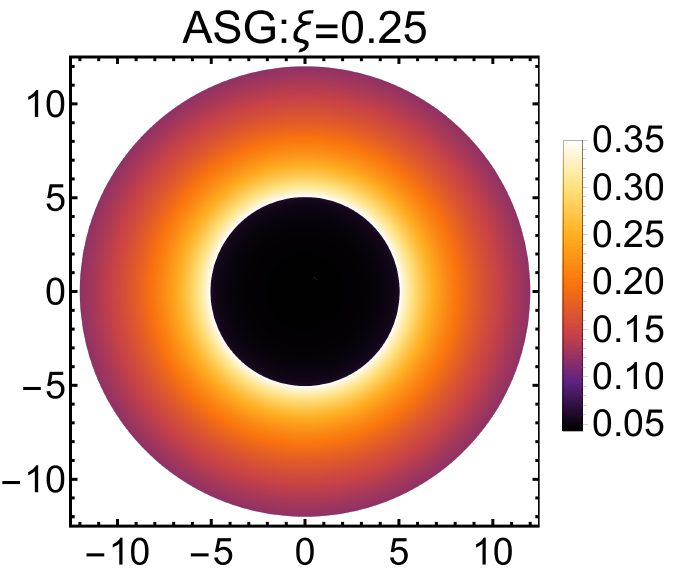}
\includegraphics[scale=0.48]{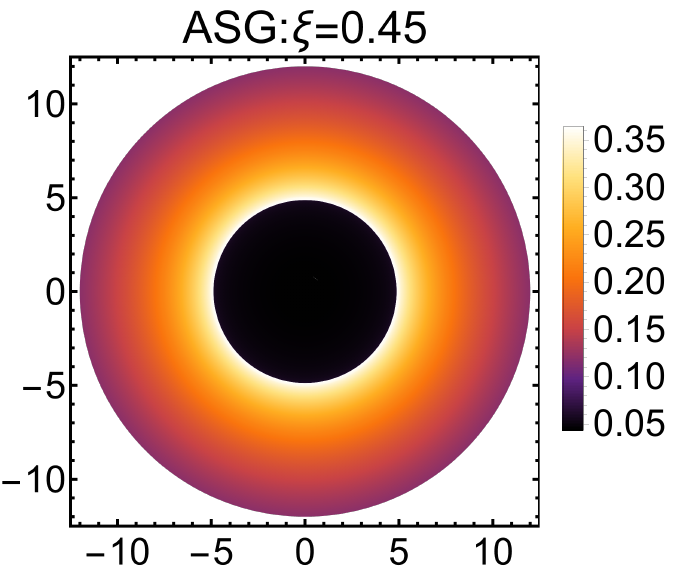}
\caption{Two-dimensional images of shadows and photon rings of the black holes with an infalling spherical accretion flow for $M=1$.}
\label{infintenfig}
\end{figure}
Similarly, the total observed intensity $I^i_{obs}$ of regular black hole surrounded by an infalling accretion flow as a function of the impact parameter $b$ has been shown in figure \ref{infallingIb}. We can clearly see that the peak values of observed intensity at the $b=b_m$ increase with the increasing of parameter $\xi$, but the black hole shadow sizes decrease. In addition, these two-dimensional planes in figure \ref{infintenfig} intuitively present the brightness of photon ring around regular black hole shadow in asymptotically safe gravity is stronger in comparison with the Schwarzschild black hole case. It is obvious that the black hole shadows with state accretion flow of the Fig.\ref{stateintenfig} are brighter than those with the infalling accretion flow of the Fig.\ref{infintenfig} due to the Doppler effect, but the size of the black hole shadows does not change with the same scale parameter $\xi$. The results imply the luminosity of the black hole shadow relies on the accretion flow model, while the size of the black hole shadow depends on the spacetime geometry.

\section{Conclusions}
\label{section5}
In this paper, we have first briefly made a review on the static spherically symmetric regular black hole metric, which is the static exterior solution of a non-singular collapsing dust ball in the context of asymptotically safe gravity. Then following the strong field limit method introduced by Bozza \cite{Bozza:2002zj} and improved by Tsukamoto \cite{Tsukamoto:2016jzh}, we have derived the analytical expression of the light deflection angle by calculating the strong field limit coefficients of the regular black hole. Meanwhile, we have evaluated the lensing observables of the supermassive black holes in this model framework as well. We finally investigated the influences of the scale parameter $\xi$ on the optical appearances of this regular black hole illuminated by different spherical accretion flows.

By the effect analysis of the scale parameter $\xi$ on the strong deflection angle and the lensing observables in regular black hole spacetime within asymptotically safe gravity, we found that the photon sphere $r_m$ and the critical impact parameter $b_m$ together decrease with the increasing of the $\xi$, while the $\bar{a}$ grows up and the $\bar{b}$ descends for the strong field limit coefficients. It also found that the strong deflection angle $\hat{\alpha}(b)$ decreases when the $\xi$ increases for the same impact parameter $b$. In addition, by supposing the regular black hole within asymptotically safe gravity as the candidate of $M87^*$ and $SgrA^*$ supermassive black holes, respectively, we saw that as the value of the $\xi$ increases, the angular positions $\vartheta_n$ and the relative magnifications $\mu_n$ of relativistic images are decreasing. The angular positions $\vartheta_n$ of relativistic images with the scale parameter $\xi$ are smaller than those of Schwarzschild black hole, and the relative magnification $\mu_n$ has the similar result for the first-order image as well. In addition, when the $\xi$ gradually increases, the position of the innermost image $\vartheta_{\infty}$ and the relative magnification $r_{mag}$ of the outermost image continuously descend, but the angular separation $s$ between the outermost and asymptotic relativistic images monotonously increases. For the above cases, these results in $SgrA^*$ are larger than those of the $M87^*$ with the same parameters. However, we found that the time delays between the first image and the second image for the $M87^*$ are much longer than those of the $SgrA^*$. Finally, by investigating the photon rings and images around regular black hole in asymptotically safe gravity, we have found that as a larger $\xi$, the photon ring for the static spherical accretion has the weaker brightness with the wider range, while the brightness of the photon ring for the radially free-infalling spherical accretion is stronger. For the same parameter $\xi$, the intensities of the black hole images in the infalling spherical accretion are weaker than those in the static
case, which are due to the Doppler effect. Meanwhile, the shadow radiuses are always the same for two types spherical accretions for the same value of $\xi$, which implies that the shadow size is only dependent of the spacetime geometry. These findings presented in this work can offer to the theoretical foundation to distinguish these non-singular black holes from their classical singular counterparts.

\acknowledgements
%We thank a lot for the helpful discussions with Profs. Ya-Peng Hu and Jun-Jin Peng, Drs. Zonghai Li.
This work was supported by the National Natural Science Foundation of China under Grant Nos. 12565009.

%\appendix
%\begin{widetext}
%\section{The motion equation of charged particles in Kerr-Newman spacetime} \label{liuhaotian}
%
%
%\section{The finite-distance deflection angle of charged particles in Kerr-Newman spacetime} \label{finidis}
%
%\end{widetext}


\begin{thebibliography}{99}
\bibitem{Einstein1916}
A. Einstein, Die Grundlage der allgemeinen Relativitatstheorie, Ann. Phys. 49 (1916) 769-822.

%\cite{LIGOScientific:2016aoc}
\bibitem{LIGO:2016aoc}
B.~P.~Abbott \textit{et al.} [LIGO Scientific and Virgo],
Observation of Gravitational Waves from a Binary Black Hole Merger,
Phys. Rev. Lett. \textbf{116}, no.6, 061102 (2016).
%doi:10.1103/PhysRevLett.116.061102
%[arXiv:1602.03837 [gr-qc]].
%11913 citations counted in INSPIRE as of 11 Nov 2024

%\cite{EventHorizonTelescope:2019dse}
\bibitem{EHT:2019dse}
K.~Akiyama \textit{et al.} [Event Horizon Telescope],
First M87 Event Horizon Telescope Results. I. The Shadow of the Supermassive Black Hole,
Astrophys. J. Lett. \textbf{875}, L1 (2019).
%doi:10.3847/2041-8213/ab0ec7
%[arXiv:1906.11238 [astro-ph.GA]].
%3375 citations counted in INSPIRE as of 11 Nov 2024

%\cite{EventHorizonTelescope:2019ths}
\bibitem{EHT:2019ths}
K.~Akiyama \textit{et al.} [Event Horizon Telescope],
First M87 Event Horizon Telescope Results. IV. Imaging the Central Supermassive Black Hole,
Astrophys. J. Lett. \textbf{875}, no.1, L4 (2019).
%doi:10.3847/2041-8213/ab0e85
%[arXiv:1906.11241 [astro-ph.GA]].
%1165 citations counted in INSPIRE as of 11 Nov 2024

%\cite{EventHorizonTelescope:2019ggy}
\bibitem{EHT:2019ggy}
K.~Akiyama \textit{et al.} [Event Horizon Telescope],
First M87 Event Horizon Telescope Results. VI. The Shadow and Mass of the Central Black Hole,
Astrophys. J. Lett. \textbf{875}, no.1, L6 (2019).
%doi:10.3847/2041-8213/ab1141
%[arXiv:1906.11243 [astro-ph.GA]].
%1323 citations counted in INSPIRE as of 11 Nov 2024

%\cite{EventHorizonTelescope:2022wkp}
\bibitem{EHT:2022wkp}
K.~Akiyama \textit{et al.} [Event Horizon Telescope], First Sagittarius A* Event Horizon Telescope Results. I. The Shadow of the Supermassive Black Hole in the Center of the Milky Way,
Astrophys. J. Lett. \textbf{930}, no.2, L12 (2022).
%doi:10.3847/2041-8213/ac6674
%406 citations counted in INSPIRE as of 06 May 2023

%\cite{EventHorizonTelescope:2022wok}
\bibitem{EHT:2022wok}
K.~Akiyama \textit{et al.} [Event Horizon Telescope], First Sagittarius A* Event Horizon Telescope Results. III. Imaging of the Galactic Center Supermassive Black Hole,
Astrophys. J. Lett. \textbf{930}, no.2, L14 (2022).
%doi:10.3847/2041-8213/ac6429
%124 citations counted in INSPIRE as of 06 May 2023

%\cite{Penrose:1964wq}
\bibitem{Penrose:1964wq}
R.~Penrose,
Gravitational collapse and space-time singularities,
Phys. Rev. Lett. \textbf{14}, 57-59 (1965).
%doi:10.1103/PhysRevLett.14.57
%1957 citations counted in INSPIRE as of 11 Nov 2024

%\cite{Hawking:1970zqf}
\bibitem{Hawking:1970zqf}
S.~W.~Hawking and R.~Penrose,
The Singularities of gravitational collapse and cosmology,
Proc. Roy. Soc. Lond. A \textbf{314}, 529-548 (1970).
%doi:10.1098/rspa.1970.0021
%1315 citations counted in INSPIRE as of 11 Nov 2024

%\cite{Penrose:1969pc}
\bibitem{Penrose:1969pc}
R.~Penrose,
Gravitational collapse: The role of general relativity,
Riv. Nuovo Cim. \textbf{1}, 252-276 (1969).
%doi:10.1023/A:1016578408204
%1774 citations counted in INSPIRE as of 13 Nov 2024

%\cite{Lan:2023cvz}
\bibitem{Lan:2023cvz}
C.~Lan, H.~Yang, Y.~Guo and Y.~G.~Miao,
Regular Black Holes: A Short Topic Review,
Int. J. Theor. Phys. \textbf{62}, no.9, 202 (2023).
%doi:10.1007/s10773-023-05454-1
%[arXiv:2303.11696 [gr-qc]].
%70 citations counted in INSPIRE as of 11 Nov 2024


%\cite{Frolov:1989pf}
\bibitem{Frolov:1989pf}
V.~P.~Frolov, M.~A.~Markov and V.~F.~Mukhanov,
THROUGH A BLACK HOLE INTO A NEW UNIVERSE?,
Phys. Lett. B \textbf{216}, 272-276 (1989).
%doi:10.1016/0370-2693(89)91114-3
%215 citations counted in INSPIRE as of 12 Nov 2024

\bibitem{Bardeen1968}
J. Bardeen, Non-singular general relativistic gravitational collapse, in Proceedings of the 5th International
Conference on Gravitation and the Theory of Relativity
(1968) p. 87.

%\cite{Ayon-Beato:2000mjt}
\bibitem{Ayon-Beato:2000mjt}
E.~Ayon-Beato and A.~Garcia,
The Bardeen model as a nonlinear magnetic monopole,
Phys. Lett. B \textbf{493}, 149-152 (2000).
%doi:10.1016/S0370-2693(00)01125-4
%[arXiv:gr-qc/0009077 [gr-qc]].
%637 citations counted in INSPIRE as of 12 Nov 2024


%\cite{Hayward:2005gi}
\bibitem{Hayward:2005gi}
S.~A.~Hayward,
Formation and evaporation of regular black holes,
Phys. Rev. Lett. \textbf{96}, 031103 (2006).
%doi:10.1103/PhysRevLett.96.031103
%[arXiv:gr-qc/0506126 [gr-qc]].
%1063 citations counted in INSPIRE as of 12 Nov 2024

%\cite{Bronnikov:2005gm}
\bibitem{Bronnikov:2005gm}
K.~A.~Bronnikov and J.~C.~Fabris,
Regular phantom black holes,
Phys. Rev. Lett. \textbf{96}, 251101 (2006).
%doi:10.1103/PhysRevLett.96.251101
%[arXiv:gr-qc/0511109 [gr-qc]].
%343 citations counted in INSPIRE as of 12 Nov 2024

%\cite{Burinskii:2001bq}
\bibitem{Burinskii:2001bq}
A.~Burinskii, E.~Elizalde, S.~R.~Hildebrandt and G.~Magli,
Regular sources of the Kerr-Schild class for rotating and nonrotating black hole solutions,
Phys. Rev. D \textbf{65}, 064039 (2002).
%doi:10.1103/PhysRevD.65.064039
%[arXiv:gr-qc/0109085 [gr-qc]].
%90 citations counted in INSPIRE as of 12 Nov 2024

%\cite{Fan:2016hvf}
\bibitem{Fan:2016hvf}
Z.~Y.~Fan and X.~Wang,
Construction of Regular Black Holes in General Relativity,
Phys. Rev. D \textbf{94}, no.12, 124027 (2016).
%doi:10.1103/PhysRevD.94.124027
%[arXiv:1610.02636 [gr-qc]].
%292 citations counted in INSPIRE as of 12 Nov 2024

%\cite{Ovalle:2023ref}
\bibitem{Ovalle:2023ref}
J.~Ovalle, R.~Casadio and A.~Giusti,
Regular hairy black holes through Minkowski deformation,
Phys. Lett. B \textbf{844}, 138085 (2023).
%doi:10.1016/j.physletb.2023.138085
%[arXiv:2304.03263 [gr-qc]].
%18 citations counted in INSPIRE as of 12 Nov 2024

%\cite{Mazza:2023iwv}
\bibitem{Mazza:2023iwv}
J.~Mazza and S.~Liberati,
Regular black holes and horizonless ultra-compact objects in Lorentz-violating gravity,
JHEP \textbf{03}, 199 (2023).
%doi:10.1007/JHEP03(2023)199
%[arXiv:2301.04697 [gr-qc]].
%9 citations counted in INSPIRE as of 12 Nov 2024

%\cite{Modesto:2010rv}
\bibitem{Modesto:2010rv}
L.~Modesto and P.~Nicolini,
Charged rotating noncommutative black holes,
Phys. Rev. D \textbf{82}, 104035 (2010).
%doi:10.1103/PhysRevD.82.104035
%[arXiv:1005.5605 [gr-qc]].
%186 citations counted in INSPIRE as of 12 Nov 2024

%\cite{Casadio:2023iqt}
\bibitem{Casadio:2023iqt}
R.~Casadio, A.~Giusti and J.~Ovalle,
Quantum rotating black holes,
JHEP \textbf{05}, 118 (2023).
%doi:10.1007/JHEP05(2023)118
%[arXiv:2303.02713 [gr-qc]].
%27 citations counted in INSPIRE as of 12 Nov 2024

%\cite{Lewandowski:2022zce}
\bibitem{Lewandowski:2022zce}
J.~Lewandowski, Y.~Ma, J.~Yang and C.~Zhang,
Quantum Oppenheimer-Snyder and Swiss Cheese Models,
Phys. Rev. Lett. \textbf{130}, no.10, 101501 (2023).
%doi:10.1103/PhysRevLett.130.101501
%[arXiv:2210.02253 [gr-qc]].
%76 citations counted in INSPIRE as of 12 Nov 2024

%\cite{Carballo-Rubio:2019fnb}
\bibitem{Carballo-Rubio:2019fnb}
R.~Carballo-Rubio, F.~Di Filippo, S.~Liberati and M.~Visser,
Geodesically complete black holes,
Phys. Rev. D \textbf{101}, 084047 (2020).
%doi:10.1103/PhysRevD.101.084047
%[arXiv:1911.11200 [gr-qc]].
%117 citations counted in INSPIRE as of 12 Nov 2024

%\cite{Bonanno:2000ep}
\bibitem{Bonanno:2000ep}
A.~Bonanno and M.~Reuter,
Renormalization group improved black hole space-times,
Phys. Rev. D \textbf{62}, 043008 (2000).
%doi:10.1103/PhysRevD.62.043008
%[arXiv:hep-th/0002196 [hep-th]].
%527 citations counted in INSPIRE as of 12 Nov 2024

%\cite{Torres:2017ygl}
\bibitem{Torres:2017ygl}
R.~Torres,
Nonsingular black holes, the cosmological constant, and asymptotic safety,
Phys. Rev. D \textbf{95}, no.12, 124004 (2017).
%doi:10.1103/PhysRevD.95.124004
%[arXiv:1703.09997 [gr-qc]].
%30 citations counted in INSPIRE as of 12 Nov 2024

%\cite{Eichhorn:2012va}
\bibitem{Eichhorn:2012va}
A.~Eichhorn,
Quantum-gravity-induced matter self-interactions in the asymptotic-safety scenario,
Phys. Rev. D \textbf{86}, 105021 (2012).
%doi:10.1103/PhysRevD.86.105021
%[arXiv:1204.0965 [gr-qc]].
%125 citations counted in INSPIRE as of 12 Nov 2024

%\cite{Pawlowski:2023dda}
\bibitem{Pawlowski:2023dda}
J.~M.~Pawlowski and J.~Tr\"ankle,
Effective action and black hole solutions in asymptotically safe quantum gravity,
Phys. Rev. D \textbf{110}, no.8, 086011 (2024).
%doi:10.1103/PhysRevD.110.086011
%[arXiv:2309.17043 [hep-th]].
%9 citations counted in INSPIRE as of 12 Nov 2024

%\cite{Stashko:2024wuq}
\bibitem{Stashko:2024wuq}
O.~Stashko,
Quasinormal modes and gray-body factors of regular black holes in asymptotically safe gravity,
Phys. Rev. D \textbf{110}, no.8, 084016 (2024).
%doi:10.1103/PhysRevD.110.084016
%[arXiv:2407.07892 [gr-qc]].
%13 citations counted in INSPIRE as of 12 Nov 2024

%\cite{Spina:2024npx}
\bibitem{Spina:2024npx}
A.~Spina, S.~Silveravalle and A.~Bonanno,
Scalar Perturbations of Regular Black Holes derived from a Non-Singular Collapse Model in Asymptotic Safety,
[arXiv:2410.05936 [gr-qc]].
%0 citations counted in INSPIRE as of 12 Nov 2024

%\cite{Bonanno:2023rzk}
\bibitem{Bonanno:2023rzk}
A.~Bonanno, D.~Malafarina and A.~Panassiti,
Dust Collapse in Asymptotic Safety: A Path to Regular Black Holes,
Phys. Rev. Lett. \textbf{132}, no.3, 031401 (2024).
%doi:10.1103/PhysRevLett.132.031401
%[arXiv:2308.10890 [gr-qc]].
%21 citations counted in INSPIRE as of 12 Nov 2024

%\cite{Markov:1985py}
\bibitem{Markov:1985py}
M.~A.~Markov and V.~F.~Mukhanov,
De sitter like initial state of the universe as a result of asymptotic disappearance of gravitational interactions of matter,
Nuovo Cim. B \textbf{86}, 97-102 (1985).
%doi:10.1007/BF02732276
%22 citations counted in INSPIRE as of 12 Nov 2024


%\cite{Bhattacharjee:2025xcb}
\bibitem{Bhattacharjee:2025xcb}
C.~Bhattacharjee, S.~Sau and A.~Mukherjee,
Radiative and jet signatures of regular black holes in quantum-corrected gravity,
Eur. Phys. J. C \textbf{85}, no.9, 1071 (2025).
%doi:10.1140/epjc/s10052-025-14725-6
%[arXiv:2509.26366 [gr-qc]].
%0 citations counted in INSPIRE as of 30 Oct 2025

%\cite{Mustafa:2025cou}
\bibitem{Mustafa:2025cou}
G.~Mustafa, A.~Alimova, F.~Atamurotov, A.~A.~Ibraheem, P.~Channuie and G.~Bahaddinova,
Testing regular black holes in the framework of asymptotically safe gravity using particle dynamics, QPOs, and shadow constraints,
Eur. Phys. J. C \textbf{85}, no.7, 741 (2025).
%doi:10.1140/epjc/s10052-025-14431-3
%7 citations counted in INSPIRE as of 30 Oct 2025


%\cite{Mannobova:2025uqf}
\bibitem{Mannobova:2025uqf}
S.~Mannobova, F.~Atamurotov, A.~Abdujabbarov, B.~S.~Alkahtani and G.~Mustafa,
Spinning particle motion around asymptotically safe gravity exhibiting regular black holes,
Eur. Phys. J. C \textbf{85}, no.5, 586 (2025).
%doi:10.1140/epjc/s10052-025-14305-8
%1 citations counted in INSPIRE as of 30 Oct 2025

%\cite{Urmanov:2025nou}
\bibitem{Urmanov:2025nou}
A.~Urmanov, H.~Chakrabarty and D.~Malafarina,
Observational properties of regular black holes in asymptotic safety,
Eur. Phys. J. C \textbf{85}, no.6, 642 (2025).
%doi:10.1140/epjc/s10052-025-14377-6
%[arXiv:2504.12072 [gr-qc]].
%5 citations counted in INSPIRE as of 30 Oct 2025

%\cite{Zhao:2025sck}
\bibitem{Zhao:2025sck}
L.~Zhao, M.~Tang and Z.~Xu,
Constraints on the scale parameter of regular black hole in asymptotically safe gravity from extreme mass ratio inspirals,
JCAP \textbf{10}, 002 (2025).
%doi:10.1088/1475-7516/2025/10/002
%[arXiv:2503.06503 [gr-qc]].
%1 citations counted in INSPIRE as of 30 Oct 2025

%\cite{Turakhonov:2025ojy}
\bibitem{Turakhonov:2025ojy}
Z.~Turakhonov, F.~Atamurotov, S.~G.~Ghosh and A.~Abdujabbarov,
Probing effects of plasma on shadow and weak gravitational lensing by regular black holes in asymptotically safe gravity,
Phys. Dark Univ. \textbf{48}, 101880 (2025).
%doi:10.1016/j.dark.2025.101880
%8 citations counted in INSPIRE as of 30 Oct 2025


\bibitem{Darwin1961}
C. Darwin, The Gravity Field of a Particle, Proc. R. Soc. London A \textbf{249}, 180 (1959); 263, 39 (1961).

%\cite{Virbhadra:1999nm}
\bibitem{Virbhadra:1999nm}
K.~S.~Virbhadra and G.~F.~R.~Ellis,
Schwarzschild black hole lensing,
Phys. Rev. D \textbf{62}, 084003 (2000).
%doi:10.1103/PhysRevD.62.084003
%[arXiv:astro-ph/9904193 [astro-ph]].
%888 citations counted in INSPIRE as of 13 Nov 2024

%\cite{Virbhadra:2002ju}
\bibitem{Virbhadra:2002ju}
K.~S.~Virbhadra and G.~F.~R.~Ellis,
Gravitational lensing by naked singularities,
Phys. Rev. D \textbf{65}, 103004 (2002).
%doi:10.1103/PhysRevD.65.103004
%643 citations counted in INSPIRE as of 13 Nov 2024

%\cite{Frittelli:1999yf}
\bibitem{Frittelli:1999yf}
S.~Frittelli, T.~P.~Kling and E.~T.~Newman,
Space-time perspective of Schwarzschild lensing,
Phys. Rev. D \textbf{61}, 064021 (2000).
%doi:10.1103/PhysRevD.61.064021
%[arXiv:gr-qc/0001037 [gr-qc]].
%238 citations counted in INSPIRE as of 13 Nov 2024

%\cite{Bozza:2002zj}
\bibitem{Bozza:2002zj}
V.~Bozza,
Gravitational lensing in the strong field limit,
Phys. Rev. D \textbf{66}, 103001 (2002).
%doi:10.1103/PhysRevD.66.103001
%[arXiv:gr-qc/0208075 [gr-qc]].
%555 citations counted in INSPIRE as of 13 Nov 2024

%\cite{Tsukamoto:2016jzh}
\bibitem{Tsukamoto:2016jzh}
N.~Tsukamoto,
Deflection angle in the strong deflection limit in a general asymptotically flat, static, spherically symmetric spacetime,
Phys. Rev. D \textbf{95}, no.6, 064035 (2017).
%doi:10.1103/PhysRevD.95.064035
%[arXiv:1612.08251 [gr-qc]].
%125 citations counted in INSPIRE as of 13 Nov 2024

%\cite{Zhang:2024sgs}
\bibitem{Zhang:2024sgs}
J.~Zhang and Y.~Xie,
Strong deflection gravitational lensing by the marginally unstable photon spheres of a wormhole,
Phys. Rev. D \textbf{109}, no.4, 043032 (2024).
%doi:10.1103/PhysRevD.109.043032
%3 citations counted in INSPIRE as of 13 Nov 2024

%\cite{Wang:2024iwt}
\bibitem{Wang:2024iwt}
Y.~Wang, A.~Vachher, Q.~Wu, T.~Zhu and S.~G.~Ghosh,
Strong Gravitational Lensing by Static Black Holes in Effective Quantum Gravity,
[arXiv:2410.12382 [astro-ph.CO]].
%0 citations counted in INSPIRE as of 13 Nov 2024

%\cite{Gao:2022gbn}
\bibitem{Gao:2022gbn}
Y.~X.~Gao and Y.~Xie,
Strong deflection gravitational lensing by an Einstein\textendash{}Lovelock ultracompact object,
Eur. Phys. J. C \textbf{82}, no.2, 162 (2022).
%doi:10.1140/epjc/s10052-022-10128-z
%17 citations counted in INSPIRE as of 13 Nov 2024

%\cite{QiQi:2023nex}
\bibitem{QiQi:2023nex}
Q.~Qi, Y.~Meng, X.~J.~Wang and X.~M.~Kuang,
Gravitational lensing effects of black hole with conformally coupled scalar hair,
Eur. Phys. J. C \textbf{83}, no.11, 1043 (2023).
%doi:10.1140/epjc/s10052-023-12233-z
%10 citations counted in INSPIRE as of 13 Nov 2024

%\cite{Furtado:2020puz}
\bibitem{Furtado:2020puz}
C.~Furtado, J.~R.~Nascimento, A.~Y.~Petrov, P.~J.~Porf\'\i{}rio and A.~R.~Soares,
Strong gravitational lensing in a spacetime with topological charge within the Eddington-inspired Born-Infeld gravity,
Phys. Rev. D \textbf{103}, no.4, 044047 (2021).
%doi:10.1103/PhysRevD.103.044047
%[arXiv:2010.11452 [gr-qc]].
%19 citations counted in INSPIRE as of 13 Nov 2024

%\cite{Nascimento:2020ime}
\bibitem{Nascimento:2020ime}
J.~R.~Nascimento, A.~Y.~Petrov, P.~J.~Porfirio and A.~R.~Soares,
Gravitational lensing in black-bounce spacetimes,
Phys. Rev. D \textbf{102}, no.4, 044021 (2020).
%doi:10.1103/PhysRevD.102.044021
%[arXiv:2005.13096 [gr-qc]].
%75 citations counted in INSPIRE as of 13 Nov 2024

%\cite{Soares:2024rhp}
\bibitem{Soares:2024rhp}
A.~R.~Soares, R.~L.~L.~Vit\'oria and C.~F.~S.~Pereira,
Topologically charged holonomy corrected Schwarzschild black hole lensing,
Phys. Rev. D \textbf{110}, no.8, 084004 (2024).
%doi:10.1103/PhysRevD.110.084004
%[arXiv:2408.03217 [gr-qc]].
%1 citations counted in INSPIRE as of 13 Nov 2024

%\cite{Kuang:2022ojj}
\bibitem{Kuang:2022ojj}
X.~M.~Kuang, Z.~Y.~Tang, B.~Wang and A.~Wang,
Constraining a modified gravity theory in strong gravitational lensing and black hole shadow observations,
Phys. Rev. D \textbf{106}, no.6, 064012 (2022).
%doi:10.1103/PhysRevD.106.064012
%[arXiv:2206.05878 [gr-qc]].
%62 citations counted in INSPIRE as of 13 Nov 2024

%\cite{Kuang:2022xjp}
\bibitem{Kuang:2022xjp}
X.~M.~Kuang and A.~\"Ovg\"un,
Strong gravitational lensing and shadow constraint from M87* of slowly rotating Kerr-like black hole,
Annals Phys. \textbf{447}, 169147 (2022).
%doi:10.1016/j.aop.2022.169147
%[arXiv:2205.11003 [gr-qc]].
%76 citations counted in INSPIRE as of 13 Nov 2024

%\cite{Gao:2021lmo}
\bibitem{Gao:2021lmo}
X.~J.~Gao, J.~M.~Chen, H.~Zhang, Y.~Yin and Y.~P.~Hu,
Investigating strong gravitational lensing with black hole metrics modified with an additional term,''
Phys. Lett. B \textbf{822}, 136683 (2021).
%doi:10.1016/j.physletb.2021.136683
%[arXiv:2108.09409 [gr-qc]].
%12 citations counted in INSPIRE as of 13 Nov 2024

%\cite{Virbhadra:2024xpk}
\bibitem{Virbhadra:2024xpk}
K.~S.~Virbhadra,
Conservation of distortion of gravitationally lensed images,
Phys. Rev. D \textbf{109}, no.12, 124004 (2024).
%doi:10.1103/PhysRevD.109.124004
%[arXiv:2402.17190 [gr-qc]].
%15 citations counted in INSPIRE as of 13 Nov 2024

%\cite{Virbhadra:2022iiy}
\bibitem{Virbhadra:2022iiy}
K.~S.~Virbhadra,
Distortions of images of Schwarzschild lensing,
Phys. Rev. D \textbf{106}, no.6, 064038 (2022).
%doi:10.1103/PhysRevD.106.064038
%[arXiv:2204.01879 [gr-qc]].
%82 citations counted in INSPIRE as of 13 Nov 2024

%\cite{Fu:2021fxn}
\bibitem{Fu:2021fxn}
Q.~M.~Fu and X.~Zhang,
Gravitational lensing by a black hole in effective loop quantum gravity,
Phys. Rev. D \textbf{105}, no.6, 064020 (2022).
%doi:10.1103/PhysRevD.105.064020
%[arXiv:2111.07223 [gr-qc]].
%29 citations counted in INSPIRE as of 13 Nov 2024

%\cite{Zhang:2017vap}
\bibitem{Zhang:2017vap}
R.~Zhang, J.~Jing and S.~Chen,
Strong gravitational lensing for black holes with scalar charge in massive gravity,
Phys. Rev. D \textbf{95}, no.6, 064054 (2017).
%doi:10.1103/PhysRevD.95.064054
%[arXiv:1805.02330 [gr-qc]].
%28 citations counted in INSPIRE as of 13 Nov 2024

%\cite{Shipley:2019kfq}
\bibitem{Shipley:2019kfq}
J.~O.~Shipley,
Strong-field gravitational lensing by black holes,''
[arXiv:1909.04691 [gr-qc]].
%12 citations counted in INSPIRE as of 13 Nov 2024





%\cite{Synge:1966okc}
\bibitem{Synge:1966okc}
J.~L.~Synge,
The Escape of Photons from Gravitationally Intense Stars,
Mon. Not. Roy. Astron. Soc. \textbf{131}, no.3, 463-466 (1966).
%doi:10.1093/mnras/131.3.463
%494 citations counted in INSPIRE as of 11 Jan 2025

%\cite{Luminet:1979nyg}
\bibitem{Luminet:1979nyg}
J.~P.~Luminet,
Image of a spherical black hole with thin accretion disk,
Astron. Astrophys. \textbf{75}, 228-235 (1979).
%765 citations counted in INSPIRE as of 11 Jan 2025

%\cite{Falcke:1999pj}
\bibitem{Falcke:1999pj}
H.~Falcke, F.~Melia and E.~Agol,
Viewing the shadow of the black hole at the galactic center,
Astrophys. J. Lett. \textbf{528}, L13 (2000).
%doi:10.1086/312423
%[arXiv:astro-ph/9912263 [astro-ph]].
%849 citations counted in INSPIRE as of 11 Jan 2025

%\cite{Cunha:2019hzj}
\bibitem{Cunha:2019hzj}
P.~Cunha, V.P., N.~A.~Eir\'o, C.~A.~R.~Herdeiro and J.~P.~S.~Lemos,
Lensing and shadow of a black hole surrounded by a heavy accretion disk,
JCAP \textbf{03}, 035 (2020).
%doi:10.1088/1475-7516/2020/03/035
%[arXiv:1912.08833 [gr-qc]].
%85 citations counted in INSPIRE as of 11 Jan 2025

%\cite{Gralla:2019xty}
\bibitem{Gralla:2019xty}
S.~E.~Gralla, D.~E.~Holz and R.~M.~Wald,
Black Hole Shadows, Photon Rings, and Lensing Rings,
Phys. Rev. D \textbf{100}, no.2, 024018 (2019).
%doi:10.1103/PhysRevD.100.024018
%[arXiv:1906.00873 [astro-ph.HE]].
%427 citations counted in INSPIRE as of 11 Jan 2025

%\cite{Narayan:2019imo}
\bibitem{Narayan:2019imo}
R.~Narayan, M.~D.~Johnson and C.~F.~Gammie,
The Shadow of a Spherically Accreting Black Hole,
Astrophys. J. Lett. \textbf{885}, no.2, L33 (2019).
%doi:10.3847/2041-8213/ab518c
%[arXiv:1910.02957 [astro-ph.HE]].
%187 citations counted in INSPIRE as of 11 Jan 2025

%\cite{Zeng:2020dco}
\bibitem{Zeng:2020dco}
X.~X.~Zeng, H.~Q.~Zhang and H.~Zhang,
Shadows and photon spheres with spherical accretions in the four-dimensional Gauss\textendash{}Bonnet black hole,
Eur. Phys. J. C \textbf{80}, no.9, 872 (2020).
%doi:10.1140/epjc/s10052-020-08449-y
%[arXiv:2004.12074 [gr-qc]].
%171 citations counted in INSPIRE as of 11 Jan 2025

%\cite{Zeng:2020vsj}
\bibitem{Zeng:2020vsj}
X.~X.~Zeng and H.~Q.~Zhang,
Influence of quintessence dark energy on the shadow of black hole,
Eur. Phys. J. C \textbf{80}, no.11, 1058 (2020).
%doi:10.1140/epjc/s10052-020-08656-7
%[arXiv:2007.06333 [gr-qc]].
%130 citations counted in INSPIRE as of 11 Jan 2025

%\cite{Zeng:2022fdm}
\bibitem{Zeng:2022fdm}
X.~X.~Zeng, M.~I.~Aslam and R.~Saleem,
The optical appearance of charged four-dimensional Gauss\textendash{}Bonnet black hole with strings cloud and non-commutative geometry surrounded by various accretions profiles,
Eur. Phys. J. C \textbf{83}, no.2, 129 (2023).
%doi:10.1140/epjc/s10052-023-11274-8
%[arXiv:2208.06246 [gr-qc]].
%28 citations counted in INSPIRE as of 08 Jan 2025

%\cite{Zeng:2024ptv}
\bibitem{Zeng:2024ptv}
X.~X.~Zeng, L.~F.~Li, P.~Li, B.~Liang and P.~Xu,
Holographic images of a charged black hole in Lorentz symmetry breaking massive gravity,
Sci. China Phys. Mech. Astron. \textbf{68}, no.2, 220412 (2025).
%doi:10.1007/s11433-024-2526-4
%[arXiv:2411.12528 [gr-qc]].
%0 citations counted in INSPIRE as of 08 Jan 2025

%\cite{Gao:2023mjb}
\bibitem{Gao:2023mjb}
X.~J.~Gao, T.~T.~Sui, X.~X.~Zeng, Y.~S.~An and Y.~P.~Hu,
Investigating shadow images and rings of the charged Horndeski black hole illuminated by various thin accretions,
Eur. Phys. J. C \textbf{83}, 1052 (2023).
%doi:10.1140/epjc/s10052-023-12231-1
%[arXiv:2311.11780 [gr-qc]].
%18 citations counted in INSPIRE as of 13 Nov 2024


%\cite{Zeng:2022pvb}
\bibitem{Zeng:2022pvb}
X.~X.~Zeng, K.~J.~He, G.~P.~Li, E.~W.~Liang and S.~Guo, QED and accretion flow models effect on optical appearance of Euler\textendash{}Heisenberg black holes,
Eur. Phys. J. C \textbf{82}, no.8, 764 (2022)
%doi:10.1140/epjc/s10052-022-10733-y
%[arXiv:2209.05938 [gr-qc]].
%6 citations counted in INSPIRE as of 10 May 2023

%\cite{Hu:2022lek}
\bibitem{Hu:2022lek}
S.~Hu, C.~Deng, D.~Li, X.~Wu and E.~Liang, Observational signatures of Schwarzschild-MOG black holes in scalar-tensor-vector gravity: shadows and rings with different accretions, Eur. Phys. J. C \textbf{82}, no.10, 885 (2022)
%doi:10.1140/epjc/s10052-022-10868-y
%10 citations counted in INSPIRE as of 10 May 2023

%\cite{Saleem:2023pyx}
\bibitem{Saleem:2023pyx}
R.~Saleem and M.~I.~Aslam,
Observable features of charged Kiselev black hole with non-commutative geometry under various accretion flow,
Eur. Phys. J. C \textbf{83}, no.3, 257 (2023)
%doi:10.1140/epjc/s10052-023-11418-w
%0 citations counted in INSPIRE as of 10 May 2023

%\cite{Zeng:2021mok}
\bibitem{Zeng:2021mok}
X.~X.~Zeng, K.~J.~He and G.~P.~Li,
Effects of dark matter on shadows and rings of Brane-World black holes illuminated by various accretions,
Sci. China Phys. Mech. Astron. \textbf{65}, no.9, 290411 (2022)
%doi:10.1007/s11433-022-1896-0
%[arXiv:2111.05090 [gr-qc]].
%22 citations counted in INSPIRE as of 10 May 2023

%\cite{Gan:2021xdl}
\bibitem{Gan:2021xdl}
Q.~Gan, P.~Wang, H.~Wu and H.~Yang,
Photon ring and observational appearance of a hairy black hole,
Phys. Rev. D \textbf{104}, no.4, 044049 (2021)
%doi:10.1103/PhysRevD.104.044049
%[arXiv:2105.11770 [gr-qc]].
%45 citations counted in INSPIRE as of 10 May 2023

%\cite{Gan:2021pwu}
\bibitem{Gan:2021pwu}
Q.~Gan, P.~Wang, H.~Wu and H.~Yang,
Photon spheres and spherical accretion image of a hairy black hole,
Phys. Rev. D \textbf{104}, no.2, 024003 (2021)
%doi:10.1103/PhysRevD.104.024003
%[arXiv:2104.08703 [gr-qc]].
%40 citations counted in INSPIRE as of 10 May 2023

%\cite{Guerrero:2021ues}
\bibitem{Guerrero:2021ues}
M.~Guerrero, G.~J.~Olmo, D.~Rubiera-Garcia and D.~S.~C.~G\'omez,
Shadows and optical appearance of black bounces illuminated by a thin accretion disk,
JCAP \textbf{08}, 036 (2021)
%doi:10.1088/1475-7516/2021/08/036
%[arXiv:2105.15073 [gr-qc]].
%51 citations counted in INSPIRE as of 10 May 2023


%\cite{Bonanno:2019ilz}
\bibitem{Bonanno:2019ilz}
A.~Bonanno, R.~Casadio and A.~Platania,
Gravitational antiscreening in stellar interiors,
JCAP \textbf{01}, 022 (2020).
%doi:10.1088/1475-7516/2020/01/022
%[arXiv:1910.11393 [gr-qc]].
%31 citations counted in INSPIRE as of 13 Nov 2024

%\cite{Bonanno:2021squ}
\bibitem{Bonanno:2021squ}
A.~Bonanno, T.~Denz, J.~M.~Pawlowski and M.~Reichert,
Reconstructing the graviton,
SciPost Phys. \textbf{12}, no.1, 001 (2022).
%doi:10.21468/SciPostPhys.12.1.001
%[arXiv:2102.02217 [hep-th]].
%87 citations counted in INSPIRE as of 13 Nov 2024


%\cite{Claudel:2000yi}
\bibitem{Claudel:2000yi}
C.~M.~Claudel, K.~S.~Virbhadra and G.~F.~R.~Ellis,
The Geometry of photon surfaces,
J. Math. Phys. \textbf{42}, 818-838 (2001).
%doi:10.1063/1.1308507
%[arXiv:gr-qc/0005050 [gr-qc]].
%491 citations counted in INSPIRE as of 13 Nov 2024


%\cite{Hu:2013eya}
\bibitem{Hu:2013eya}
Y.~P.~Hu, H.~S.~Zhang, J.~P.~Hou and L.~Z.~Tang,
Perihelion precession and deflection of light in the general spherically symmetric spacetime,
Adv. High Energy Phys. \textbf{2014}, 604321 (2014).
%doi:10.1155/2014/604231
%[arXiv:1312.7419 [gr-qc]].
%17 citations counted in INSPIRE as of 13 Nov 2024

%\cite{Gao:2019pir}
\bibitem{Gao:2019pir}
X.~Gao, S.~Song and J.~Yang,
Light bending and gravitational lensing in Brans-Dicke theory,
Phys. Lett. B \textbf{795}, 144-151 (2019).
%doi:10.1016/j.physletb.2019.06.028
%[arXiv:1905.07968 [gr-qc]].
%10 citations counted in INSPIRE as of 13 Nov 2024

%\cite{Keeton:2005jd}
\bibitem{Keeton:2005jd}
C.~R.~Keeton and A.~O.~Petters,
Formalism for testing theories of gravity using lensing by compact objects. I. Static, spherically symmetric case,
Phys. Rev. D \textbf{72}, 104006 (2005).
%doi:10.1103/PhysRevD.72.104006
%[arXiv:gr-qc/0511019 [gr-qc]].
%210 citations counted in INSPIRE as of 14 Nov 2024

%\cite{Gao:2024ejs}
\bibitem{Gao:2024ejs}
X.~J.~Gao,
Gravitational lensing and shadow by a Schwarzschild-like black hole in metric-affine bumblebee gravity,
Eur. Phys. J. C \textbf{84}, no.9, 973 (2024).
%doi:10.1140/epjc/s10052-024-13338-9
%[arXiv:2409.12531 [gr-qc]].
%1 citations counted in INSPIRE as of 13 Nov 2024

%\cite{Virbhadra:1998dy}
\bibitem{Virbhadra:1998dy}
K.~S.~Virbhadra, D.~Narasimha and S.~M.~Chitre,
Role of the scalar field in gravitational lensing,
Astron. Astrophys. \textbf{337}, 1-8 (1998).
%[arXiv:astro-ph/9801174 [astro-ph]].
%424 citations counted in INSPIRE as of 13 Nov 2024

%\cite{Bozza:2008ev}
\bibitem{Bozza:2008ev}
V.~Bozza,
A Comparison of approximate gravitational lens equations and a proposal for an improved new one,
Phys. Rev. D \textbf{78}, 103005 (2008).
%doi:10.1103/PhysRevD.78.103005
%[arXiv:0807.3872 [gr-qc]].
%151 citations counted in INSPIRE as of 13 Nov 2024

%\cite{Bozza:2003cp}
\bibitem{Bozza:2003cp}
V.~Bozza and L.~Mancini,
Time delay in black hole gravitational lensing as a distance estimator,
Gen. Rel. Grav. \textbf{36}, 435-450 (2004).
%doi:10.1023/B:GERG.0000010486.58026.4f
%[arXiv:gr-qc/0305007 [gr-qc]].
%128 citations counted in INSPIRE as of 24 Dec 2024


%\cite{Yuan:2014gma}
\bibitem{Yuan:2014gma}
F.~Yuan and R.~Narayan,
Hot Accretion Flows Around Black Holes,
Ann. Rev. Astron. Astrophys. \textbf{52}, 529-588 (2014).
%doi:10.1146/annurev-astro-082812-141003
%[arXiv:1401.0586 [astro-ph.HE]].
%1000 citations counted in INSPIRE as of 06 Feb 2025

%\cite{Jaroszynski:1997bw}
\bibitem{Jaroszynski:1997bw}
M.~Jaroszynski and A.~Kurpiewski,
Optics near kerr black holes: spectra of advection dominated accretion flows,
Astron. Astrophys. \textbf{326}, 419 (1997).
%[arXiv:astro-ph/9705044 [astro-ph]].
%68 citations counted in INSPIRE as of 17 May 2023

%\cite{Bambi:2013nla}
\bibitem{Bambi:2013nla}
C.~Bambi, Can the supermassive objects at the centers of galaxies be traversable wormholes? The first test of strong gravity for mm/sub-mm very long baseline interferometry facilities,
Phys. Rev. D \textbf{87}, 107501 (2013).
%doi:10.1103/PhysRevD.87.107501
%[arXiv:1304.5691 [gr-qc]].
%244 citations counted in INSPIRE as of 09 Feb 2025

%\cite{Wang:2023vcv}
\bibitem{Wang:2023vcv}
X.~J.~Wang, X.~M.~Kuang, Y.~Meng, B.~Wang and J.~P.~Wu,
Rings and images of Horndeski hairy black hole illuminated by various thin accretions,
Phys. Rev. D \textbf{107}, no.12, 124052 (2023).
%doi:10.1103/PhysRevD.107.124052
%[arXiv:2304.10015 [gr-qc]].
%43 citations counted in INSPIRE as of 09 Feb 2025

\end{thebibliography}
\end{document}